\documentclass[]{tMPH2e}

\usepackage[nolists]{endfloat}
\usepackage{longtable}
\DeclareDelayedFloatFlavor*{longtable}{table}
\usepackage{graphicx}
\usepackage{wasysym}
\usepackage{amssymb}
\usepackage{amsmath}
\usepackage[usenames,dvipsnames]{color}
\newcommand\solidrule[1][0.5cm]{\rule[0.5ex]{#1}{.4pt}}
\newcommand\dashedrule{\mbox{\solidrule[1mm]\hspace{1mm}\solidrule[1mm]\hspace{1mm}\solidrule[1mm]}}

\begin{document}

%\doublespacing
\doi{10.1080/0892702YYxxxxxxxx}
\issn{1029-0435}  \issnp{0892-7022}
\jvol{00} \jnum{00} \jyear{2016} %\jmonth{15 January}

\markboth{Stephan Werth, Katrin St\"obener, Martin Horsch, Hans Hasse}{Molecular Physics}

\articletype{}

\title{Simultaneous description of bulk and interfacial properties of fluids by the Mie potential}

%\author{Stephan Werth$^1$, Katrin St\"obener$^2$, Martin Horsch$^{1,\ast}$\thanks{$^\ast$Corresponding author. Email: martin.horsch@mv.uni-kl.de
%} and Hans Hasse$^1$ \\{\vbox{\vspace{12pt} {\em{$^1$ Laboratory of Engineering Thermodynamics, Department of Mechanical and Process Engineering, University of %Kaiserslautern,
%Erwin-Schr\"odinger-Str. 44, 67663 Kaiserslautern, Germany}}
%\\{\vbox{\vspace{12pt} {\em{$^2$ Fraunhofer Institute for Industrial Mathematics, Department for Optimization, Fraunhofer-Platz 1, 67663 Kaiserslautern, Germany}}\received{submitted April 2015}}} }

\author{Stephan Werth$^1$, Katrin St\"obener$^2$, Martin Horsch$^{1\ast}$\thanks{$^\ast$Corresponding author. Email: martin.horsch@mv.uni-kl.de
\vspace{6pt}} \& Hans Hasse$^1$ \\{\vbox{\vspace{12pt} {\em{$^1$Laboratory of Engineering Thermodynamics, Department of Mechanical and Process Engineering, University of Kaiserslautern,
Erwin-Schr\"odinger-Str. 44, 67663 Kaiserslautern, Germany}}
\\ {\em{$^2$Fraunhofer Institute for Industrial Mathematics, Department for Optimization, Fraunhofer-Platz 1, 67663 Kaiserslautern, Germany}}\\\vspace{6pt}\received{submitted May 2016}}} }

\maketitle

\begin{abstract}
The vapor-liquid equilibrium (VLE) of the Mie potential, where the dispersive exponent is constant ($m =6$) while the repulsive exponent $n$ is varied between 9 and 48, is systematically investigated by molecular simulation. 
For systems with planar vapor-liquid interfaces, long-range correction expressions are derived, so that interfacial and bulk properties can be computed accurately. 
The present simulation results are found to be consistent with the available body of literature on the Mie fluid which is substantially extended.
On the basis of correlations for the considered thermodynamic properties, a multicriteria optimization becomes viable. Thereby, users can adjust the three parameters of the Mie potential to the properties of real fluids, weighting different thermodynamic properties according to their importance for a particular application scenario. 
In the present work, this is demonstrated for carbon dioxide for which different competing objective functions are studied which describe the accuracy of the model for representing the saturated liquid density, the vapor pressure and the surface tension. It is shown that models can be found which describe simultaneously the saturated liquid density and vapor pressure with good accuracy, and it is discussed to what extent this accuracy can be upheld as the model accuracy for the surface tension is further improved.
\bigskip

\begin{keywords}Mie potential; surface tension; molecular simulation; long-range correction; carbon dioxide
\end{keywords}\bigskip

\end{abstract}

\section{Introduction}

In process engineering, knowledge of the vapor-liquid equilibrium is crucial for process design. Molecular modelling and simulation based on force fields is a promising way of predicting these thermodynamic properties. However, it is a topic of controversial discussion to what extent effective pair potentials are capable of reproducing bulk and interfacial properties of real fluids at the same time \cite{GPMMLB09,ORD08,WHH16,WSKKHH15,SKRMKH14,SKHKH15,ALGAJM11,ALAGMJ13,EV15,ZLCMSA13,GMT14,NWLM12,SE06,FMGLI16}. The present work reports on bulk and interfacial properties of the Mie fluid and adresses the question of the simultaneous description of these properties by that model. %Thermodynamic properties of the vapor-liquid equilibrium of the Mie fluid as a whole are determined. The present work also characterizes the level of agreement that can be achieved for CO$_2$ using a three-parameter Mie potential. 

The Mie potential \cite{M03,G12} is a generalized version of the Lennard-Jones potential \cite{L24,L31} with variable exponents for the repulsive and dispersive interactions. The Lennard-Jones potential has a dispersive exponent of 6, which is physically motivated \cite{L30}. This exponent ($m = 6$) is used here throughout. It is, however, noted that also the exponent $m$ has been varied in studies in the literature \cite{VLAAJMG14,MJ14,ALAGMJ13,LALMJ15,ALGAJM11}.
The repulsive exponent of the Lennard-Jones potential was originally set to $n=12$ for numerical reasons rather than for physical reasons. Here, therefore, $n$ is varied. Vapor-liquid equilibrium (VLE) data for the single-site Mie potential are available in the literature for many combinations of the repulsive and dispersive exponents \cite{RAMG15,OY00,N08,KSGP96,P00,ORD08,GPMMLB09}. Interfacial properties of the Mie potential were only reported so far by Orea et al. \cite{ORD08} and Galliero et al. \cite{GPMMLB09}. 

%Most common transferable force fields are based on the Lennard-Jones potential and superimposed electrostatics, e.g. OPLS \cite{JMS84}, TraPPE \cite{MS98}, AUA \cite{UBDBRF00}, NERD \cite{NED98}, AMBER \cite{WWCKC05} and GROMOS \cite{GROMOS}. 
While it is certain that due to the third parameter the Mie potential (with a fixed $m$ = 6) must be better suited for correlations of experimental data than the two parameter Lennard-Jones potential, it is not self-evident how large the improvement in model accuracy can become.% and how the value of the repulsive exponent parameter $m$ needs to be varied.

%So far, only few comprehensive studies concerning the Mie potential have been performed \cite{RAMG15,OY00,N08,KSGP96,P00,ORD08,GPMMLB09}.
%However, 
There are several transferable force field parameter sets for the Mie potential in the literature, e.g. for $n$-alkanes \cite{PB09,HG15}, perfluoroalkanes \cite{PB09}, alkenes \cite{PK14}, $n$-olefins \cite{HG15}, ethers \cite{HPG15}. For mixtures of $n$-alkanes with noble gases, Mick et al. \cite{MBJRSP15} developed force fields based on the Mie potential. %Coarse-grained models exist for alkanes \cite{MS11}.
The above mentioned models \cite{PB09,HG15,PK14,HPG15,MBJRSP15} use a dispersive exponent $m = 6$. 
A repulsive exponent $n = 16$ is used for alkenes and alkanes \cite{PK14,PB09}, $n = 14$ for olefins, alkanes and methane \cite{HG15,PB09}, $n = 12$ for ether groups \cite{HPG15}, and $36 \leq n \leq 44$ for perfluoroalkanes \cite{PB09}. For mixed interactions, the arithmetic mean value is used for the exponents \cite{PB09}. The parameters of these molecular models were adjusted to bulk properties of the VLE \cite{PB09,HG15,PK14,HPG15,MS11,MBJRSP15}. %The exponent for the dispersion is usually not varied and remains 6, due to the London forces \cite{L30}. 
Moreover, Jackson and co-workers developed a large number of coarse grained models based on the Mie potential, e.g. for CO$_2$ \cite{ALGAJM11}, CF$_4$, SF$_6$, R1234yf, $n$-C$_{10}$H$_{22}$, C$_{20}$H$_{42}$ \cite{ALAGMJ13}, benzene, $n$-decylbenzene \cite{LAPGAJM12} and water \cite{LALMJ15}. The parameterization was done indirectly, using the SAFT-$\gamma$-Mie \cite{VLAAJMG14,MJ14,ALAGMJ13} or SAFT-VR-Mie equation of state \cite{LAAGAMJ13,DLGJH15}. %Optimization of the Mie potential based on the SAFT-VR equation of state \textbf{(keine mol sim!)}, pseudo-multi-criteria optimization \cite{DLGJH15}. Grundlage SAFT VR Mie

%CO$_2$: Maurers Habil 33-7 and 30-7  \cite{M78}, Avendano 23-6.66 \cite{ALGAJM11}
%CF$_4$: 32.53 - 6 \cite{ALAGMJ13}
%SF$_6$: 19.02 - 8.8 \cite{ALAGMJ13}
%Water: 8 - 6 \cite{LALMJ15}

In the present work, the single-site three-parameter Mie potential with $m = 6$ and $9 \leq n \leq 48$ is studied systematically: the saturated liquid density, the saturated vapor density, the vapor pressure, the enthalpy of vaporization and the surface tension are determined by molecular dynamics simulation of systems that contain a vapor phase and a liquid phase (and the interface between them) and correlated as a function of the model parameters. For this purpose, a long-range correction of the Mie potential is developed for inhomogeneous simulation volumes with planar symmetry. Correlation expressions are derived for the investigated thermodynamic properties of the three-parameter Mie potential. In a case study, these correlations are used for the parameterization of a Mie potential for carbon dioxide. Multicriteria optimization is applied, taking into account several conflicting objective functions: the vapor pressure, the saturated liquid density, and the surface tension.

The work presented here for the Mie potential extends previous work of our group on other molecular model classes, namely the 2CLJQ \cite{SVHF01,WHH15,FVH06,VSH01,WSKKHH15,SKHKH15,FVH06b,FVH05,FVH205} and the 2CLJD \cite{SVH03b,WHH16,FVH06a,SVH03a} potential. %study of the Mie potential is conducted, starting from the LRC for heterogeneous systems, over the VLE data of the model fluid to molecular model optimization based on the Pareto set. Thereby, an dispersive term proportional to $r^{-6}$ is used in all cases, while the repulsive exponent is varied between 9 and 48.

\section{Simulations with the Mie potential}

The Mie potential is given by \cite{M03,G12}
\begin{equation}
 u(r) = \frac{n}{n-m} \left( \frac{n}{m}\right)^\frac{m}{n-m} \epsilon \left[ \left(\frac{\sigma}{r}\right)^n - \left(\frac{\sigma}{r}\right)^m\right],
 \label{eq:Mie_potential}
\end{equation}
where $\sigma$ and $\epsilon$ are the size and energy parameter, $n$ and $m$ are the repulsive and dispersive exponents and $r$ represents the distance between two interaction sites.%, which may deviate from the distance $r$ between the centers of mass. Fig. \ref{fig:Mie_AngleAveraging} shows the three possible cases for the interaction type. The center-site (CS) and site-site (SS) cases are discussed in the Appendix. In the following, the center-center (CC) case is discussed, where $r=s$. %For consistency with the approach by London, the dispersive exponent will be $m$ = 6 \cite{L30}.

For numerical reasons, the intermolecular pair potential in molecular simulation needs to be truncated. However, thermodynamic properties in heterogeneous systems are very sensitive to a truncation of the intermolecular potential \cite{ZLCMSA13,LVF90,GMT14,WHVH15,GMT15,GGMT15,WHH15,GMSBRF04,FMDYS15,WMPAHL15,MMB15,WSZ12,NOBK16}. For the Lennard-Jones potential, a large variety of long-range corrections (LRC) exist for heterogeneous systems to account for the inhomogeneity, ranging from Ewald summation techniques \cite{VIG07,IHMI12,IHMHKE13}, the Fast Multipole Method (FMM) \cite{MENT03} and Multilevel Summation (MLS) \cite{TSBI14} to slab-based LRC techniques \cite{MWF97,Janecek06,WRVHH14}. In terms of the thermodynamic results, the different methods deliver a similar degree of accuracy for Lennard-Jones systems \cite{Janecek06,IHMI12,TSBI14,WRVHH14}. For the Mie potential, no LRC for heterogeneous systems exist, to the best of our knowledge. 
To overcome this problem, large cutoff radii are used in the literature (up to 35 \AA \cite{LALMJ15} or 10 molecular segment diameters \cite{GPMMLB09}). Not only the dispersive exponent has an influence on the magnitude of the long-range interactions, which is obvious, but also the repulsive exponent \cite{GPMMLB09}.

The LRC by Jane\v{c}ek \cite{Janecek06} for single Lennard-Jones sites based on the density profile can be straightforwardly generalized to the Mie potential.
The LRC for the potential energy is given by
\begin{equation}
 U_i^{\rm LRC} = \sum_k^{N_{\rm s}} 2 \pi \rho(y_k) \Delta y \int_{r'}^\infty \text{d}r \frac{n}{n-m} \left( \frac{n}{m}\right)^\frac{m}{n-m} \epsilon \left[ \left( \frac{\sigma}{r}\right)^n - \left( \frac{\sigma}{r}\right)^m\right],
 \label{eq:Mie_LRC}
\end{equation}
where $\rho(y_k)$ is the density in slab $k$, $\Delta y$ is the thickness of a slab and $N_{\rm s}$ is the number of slabs. The lower bound for the integration $r'$ is defined according to Siperstein et al. \cite{SMT02}: if the distance $\xi = |y_i-y_k|$ between a molecule $i$ and the slab $k$ is smaller than the cutoff radius, the cutoff radius is used, and $\xi$ is used otherwise, i.e.
\begin{equation}
 r' =\begin{cases}
  r_{\rm c}, \quad \text{if} \quad \xi < r_{\rm c} \\
  \xi, \quad \text{else}.
 \end{cases}
 \label{eq:Mie_Cases}
\end{equation}
From Eqs.\ (\ref{eq:Mie_LRC}) and (\ref{eq:Mie_Cases}), the following expressions for the contributions of the long-range correction to the potential energy $U_i$, the force $F_i$ and the normal and tangential virial $\Pi_{\mathrm{N},i}$,$\Pi_{\mathrm{T},i}$ can be derived,
%The definition of the lower integration bound is combined with Eq. (\ref{eq:Mie_LRC}) for the potential energy and the corresponding equations for the force and the virial, i.e.
\begin{align}
U_i^{\rm LRC} &= \sum_k^{N_s} 2 \pi \rho(y_k) \Delta y \int_{r'}^\infty \text{d}r \frac{n}{n-m} \left( \frac{n}{m}\right)^\frac{m}{n-m} \epsilon \left[ \left( \frac{\sigma}{r}\right)^n - \left( \frac{\sigma}{r}\right)^m\right] r \notag \\
&= \sum_k^{N_s} 2 \pi \rho(y_k) \Delta y \frac{n}{n-m} \left( \frac{n}{m}\right)^\frac{m}{n-m} \epsilon \sigma^2 \Bigg[ \frac{1}{n-2} \left( \frac{\sigma}{r'}\right)^{n-2} \nonumber \\ & \quad \quad \quad \quad \quad \quad - \frac{1}{m-2} \left( \frac{\sigma}{r'}\right)^{m-2}\Bigg],
\end{align}

\begin{align}
F_i^{\rm LRC} &= - \sum_k^{N_s} 2 \pi \rho(y_k) \Delta y \int_{r'}^\infty \text{d}r \frac{\partial u}{\partial r} \frac{\xi}{r} r \notag \\
&= - \sum_k^{N_s} 2 \pi \rho(y_k) \Delta y \xi \frac{n}{n-m} \left( \frac{n}{m}\right)^\frac{m}{n-m} \epsilon \left[ \left( \frac{\sigma}{r'}\right)^{n} - \left( \frac{\sigma}{r'}\right)^{m}\right], \\
% \end{align}
%
% \begin{align}
 \Pi_{\mathrm{N},i}^{\rm LRC} &= \sum_k^{N_s} \pi \rho(y_k) \Delta y \int_{r'}^\infty \text{d}r \frac{\partial u}{\partial r} \frac{\xi^2}{r} r \notag \\
 &= \sum_k^{N_s} \pi \rho(y_k) \Delta y \xi ^2 \frac{n}{n-m} \left( \frac{n}{m}\right)^\frac{m}{n-m} \epsilon \left[ \left( \frac{\sigma}{r'}\right)^{n} - \left( \frac{\sigma}{r'}\right)^{m}\right], \\
%\end{align}
%
%\begin{align}
 \Pi_{\mathrm{T},i}^{\rm LRC} &= \sum_k^{N_s} \frac{1}{2} \pi \rho(y_k) \Delta y \int_{r'}^\infty \text{d}r \frac{\partial u}{\partial r} \frac{r^2 - \xi^2}{r} r \notag \\
 &= \sum_k^{N_s} \frac{1}{2} \pi \rho(y_k) \Delta y \frac{n}{n-m} \left( \frac{n}{m}\right)^\frac{m}{n-m} \epsilon \Bigg[ \frac{n r^2 - (n-2)\xi^2}{(n-2)} \left( \frac{\sigma}{r'}\right)^{n} \nonumber \\ & \quad \quad \quad \quad \quad \quad - \frac{mr^2-(m-2)\xi^2}{m-2}\left( \frac{\sigma}{r'}\right)^{m}\Bigg].
\end{align}
These correction terms are valid for single-site Mie models and a center-of-mass cutoff scheme, or for multi-site models if a site-site cutoff radius scheme is used. The corresponding expression for multi-site models and a center-of-mass cutoff scheme are given in the supplementary material.

In the present work, systems were studied where the vapor and liquid phases coexist in direct contact, employing periodic boundary conditions, so that there are two vapor-liquid interfaces which are oriented perpendicular to the $y$ axis. The surface tension was computed from the deviation between the normal and the tangential diagonal components of the overall pressure tensor \cite{WTRH83,IK49}
\begin{equation}
 \gamma = \frac{1}{2A}\left(\Pi_{\rm N} - \Pi_{\rm T}\right) = \frac{1}{2}\int_{-\infty}^\infty \text{d}y \left(p_{\rm N} - p_{\rm T} \right).
\end{equation}
Thereby, the normal pressure $p_{\rm N}$ is given by the $y$ component of the diagonal of the pressure tensor, and the tangential pressure $p_{\rm T}$ was determined by averaging over the $x$ and $z$ components of the diagonal of the pressure tensor. The surface area $A$ of each vapor-liquid interface is given by the cross section of the simulation volume normally to the $y$ axis.

All thermodynamic properties can be reduced by the Lennard-Jones parameters $\sigma$ and $\epsilon$, the mass $m$, as well as the Boltzmann constant $k_{\rm B}$. This approach reduces the parameters of the Mie fluid as it is studied here to one, the repulsive exponent $n$. Molecular simulations were performed in the present work for 14 different repulsive exponents ranging from $n = 9$ to $n = 48$ (in steps of $\Delta n = 3$). The temperature was varied from approximately 55 to 95 \% $T_{\rm c}$, where $T_{\rm c}$ is the critical temperature of the studied fluids.% estimated by a correlation based on literature data. 

The simulations were performed with an extended version of the molecular dynamics code $ls1$ $mardyn$~\cite{ls12013,WHB13} in the canonical ensemble with $N$ = 16,000 particles. Further simulation details are given in the Appendix.

\section{Simulation results}

\subsection{Long-range correction}

To validate the LRC, a series of simulations of the Mie fluid in VLE at approximately 55 \% of the critical temperature was conducted. Unless a suitable LRC is used, the critical temperature is not reproduced correctly, and permanent homogeneous configurations may be found in the simulation even in the two-phase region significantly below the actual critical temperature. % critical temperature depends on the cutoff radius, if no suitable LRC is used. 
Therefore a temperature close to the triple point temperature was used for the systematic study of the influence of the LRC, so that vapor-liquid equilibria were also obtained for the extreme case of a short cutoff radius without any LRC.

\begin{figure}[htb]
\centering
 \includegraphics[scale = 0.95]{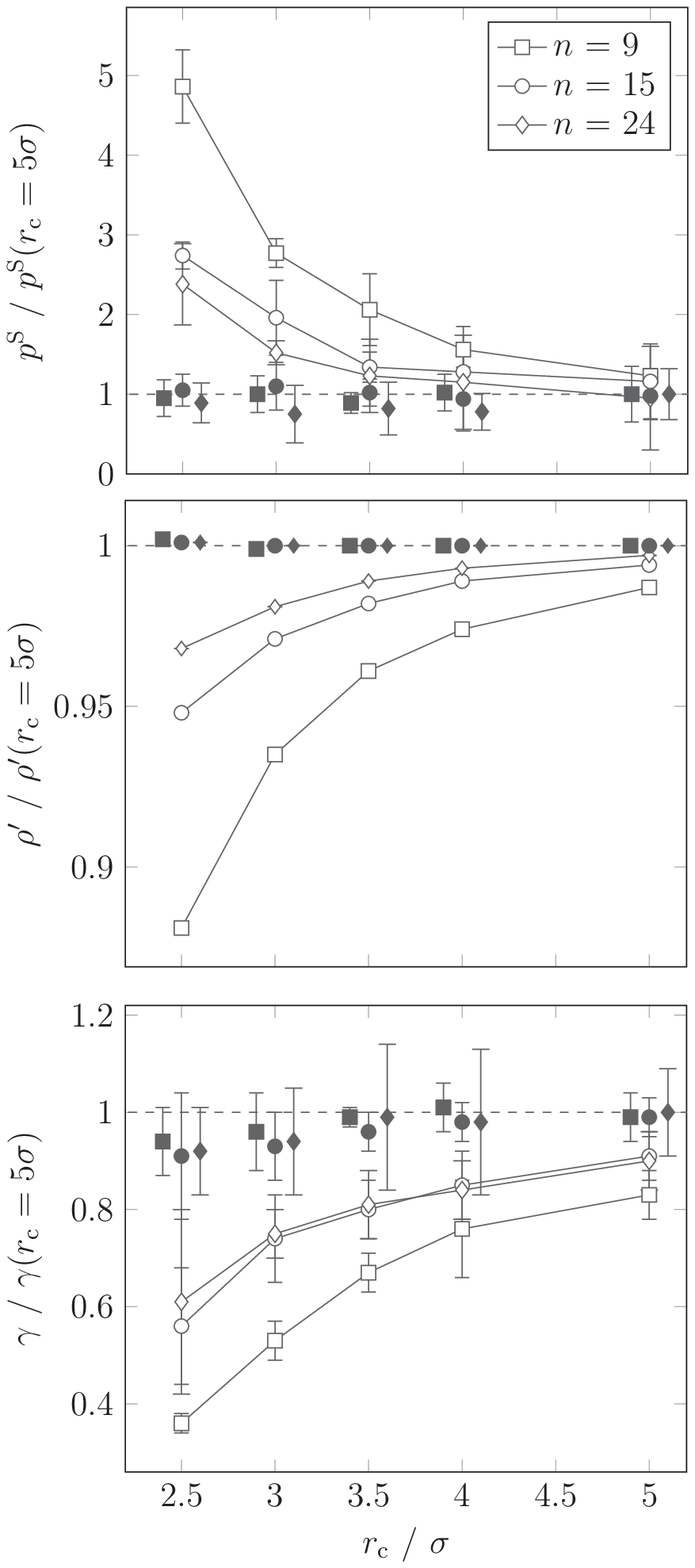}
\caption{Influence of the cutoff radius and the LRC on different thermodynamic properties of the Mie fluid at 55 \% of their critical temperature: Vapor pressure (top), saturated liquid density (center) and surface tension (bottom). These thermodynamic properties are reduced here by the values obtained for a cutoff radius of $r_{\rm c} = 5$ $\sigma$, using the LRC from the present work. The open symbols correspond to simulations without any LRC and the closed symbols are the simulation results with the present LRC. The results for the present LRC and $n = 9$ are shifted to the left by 0.1 $\sigma$, and the results for $n = 24$ are shifted to the right by 0.1 $\sigma$, to make the results clearly visible.}  
\label{fig:Mie_LRC}
\end{figure}

In Fig. \ref{fig:Mie_LRC}, these results are compared to the simulation results with a cutoff radius of $r_{\rm c} = 5$ $\sigma$ and the present LRC approach to enable a comparison of various exponents. The results show the importance of the LRC for the heterogeneous simulations carried out in the present study. Without LRC, the results depend on the chosen value for $r_{\rm c}$ very significantly, so that large cutoff radii need to be used, which makes the simulations numerically expensive. 
For all thermodynamic properties, the results without LRC do eventually converge to a limit, but this limit is not reached even for $r_{\rm c} = 5$ $\sigma$. Galliero et al. \cite{GPMMLB09} showed that the saturated liquid density, vapor pressure and surface tension converge to the correct values if a cutoff radius of at least 7 $\sigma$ is used for Mie fluids with $8 < n < 20$. In contrast, using the LRC presented above, the results depend hardly on the choice of $r_{\rm c}$, even for small values. Upon increasing $r_{\rm c}$, the results obtained without LRC converge to those obtained with LRC, but only very slowly \cite{LALMJ15,GPMMLB09}. This shows that the LRC as presented above and implemented in $ls1$ $mardyn$ is correct and efficient.
Fig. \ref{fig:Mie_LRC} also shows that the influence of the cutoff radius on the results obtained without LRC depends on the repulsive exponent used in the Mie potential, and that it is larger for small exponents.
%The simulation results without a LRC exhibit deviations of up to 380 \% in the vapor pressure, 12 \% in the saturated liquid density and 64 \% in the surface tension for the smallest cutoff radius used in the present work. For largest cutoff radii used in the present work, i.e. $r_c = 5$ $\sigma$, deviations due to neglecting long-range effects are of the order of 20 \% for the vapor pressure, 1 \% in the saturated liquid density and 10 \% in the surface tension. The simulation results with the present LRC approach exhibit hardly any dependence on the cutoff radius.

The additional time consumption for the LRC is of the order of 10 \% for a cutoff radius of $r_{\rm c} = 2.5$ $\sigma$ and becomes negligible for larger cutoff radii \cite{WRVHH14,WHH15}. In contrast to Ewald summation based techniques, the LRC based on the density profile can also be applied to large numbers of particles and processing units, due to the low amount of communication needed for the evaluation of the density profile \cite{IHMHKE13}. All simulation results reported below are obtained using the present LRC approach and a constant cutoff radius of $r_{\rm c} = 5$ $\sigma$.

\subsection{Systematic study of the vapor-liquid equilibrium}

\begin{figure}[htb]
\centering
 \includegraphics{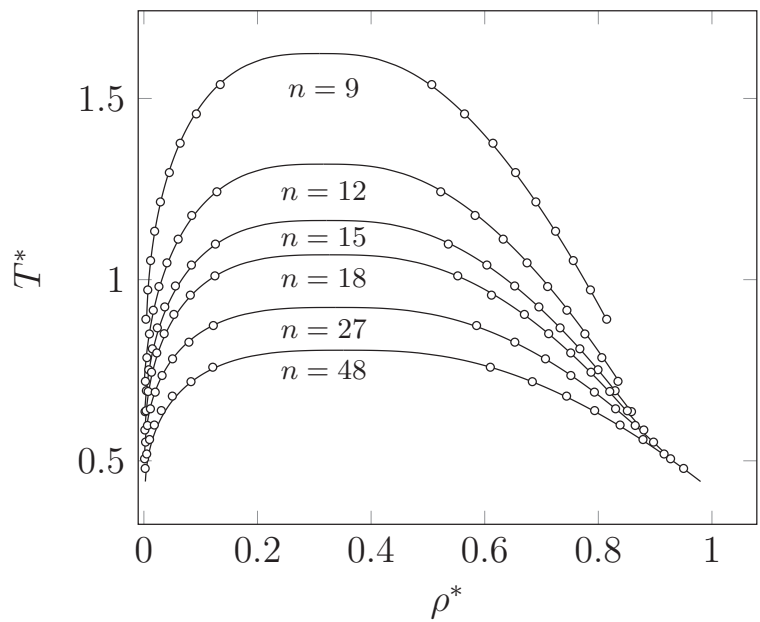}
\caption{Saturated densities of the Mie fluid. The symbols are simulation results from the present work. The lines are correlations from Eqs.\ (\ref{eq:Mie_DensityLiquid}) and (\ref{eq:Mie_DensityVapor}). The simulation uncertainties are smaller than the symbol size in all cases.}  
\label{fig:Mie_rhoT}
\end{figure}

\begin{figure}[htb]
\centering
 \includegraphics{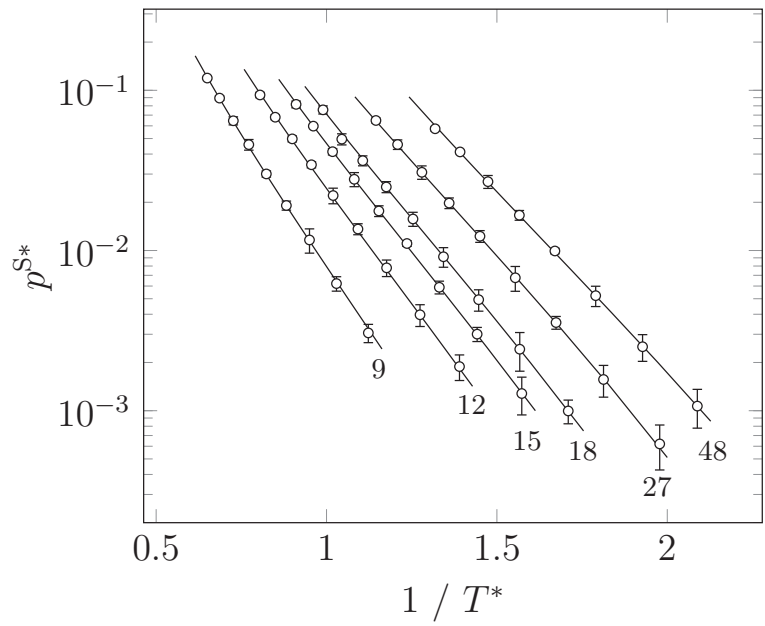}
\caption{Vapor pressure curves of the Mie fluid. The symbols are simulation results from the present work. The lines are correlations from Eq. (\ref{eq:Mie_VaporPressure}). The numbers are those for the exponent $n$ of the fluid.}  
\label{fig:Mie_pT}
\end{figure}

\begin{figure}[htb]
\centering
 \includegraphics{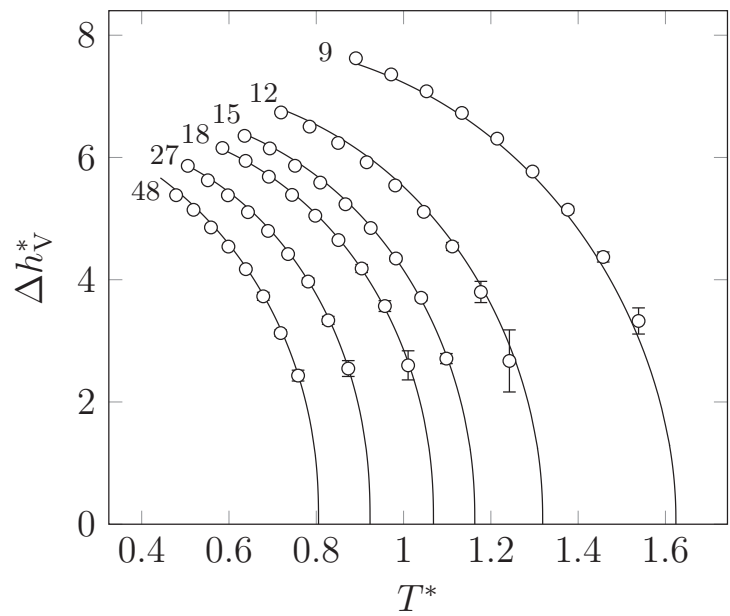}
\caption{Enthalpy of vaporization of the Mie fluid as a function of the temperature. The symbols are simulation results from the present work. The lines are correlations from Eq. (\ref{eq:Mie_Enthalpy}). The numbers are those for the exponent $n$ of the fluid.}  
\label{fig:Mie_deltahV}
\end{figure}

\begin{figure}[htb]
\centering
 \includegraphics{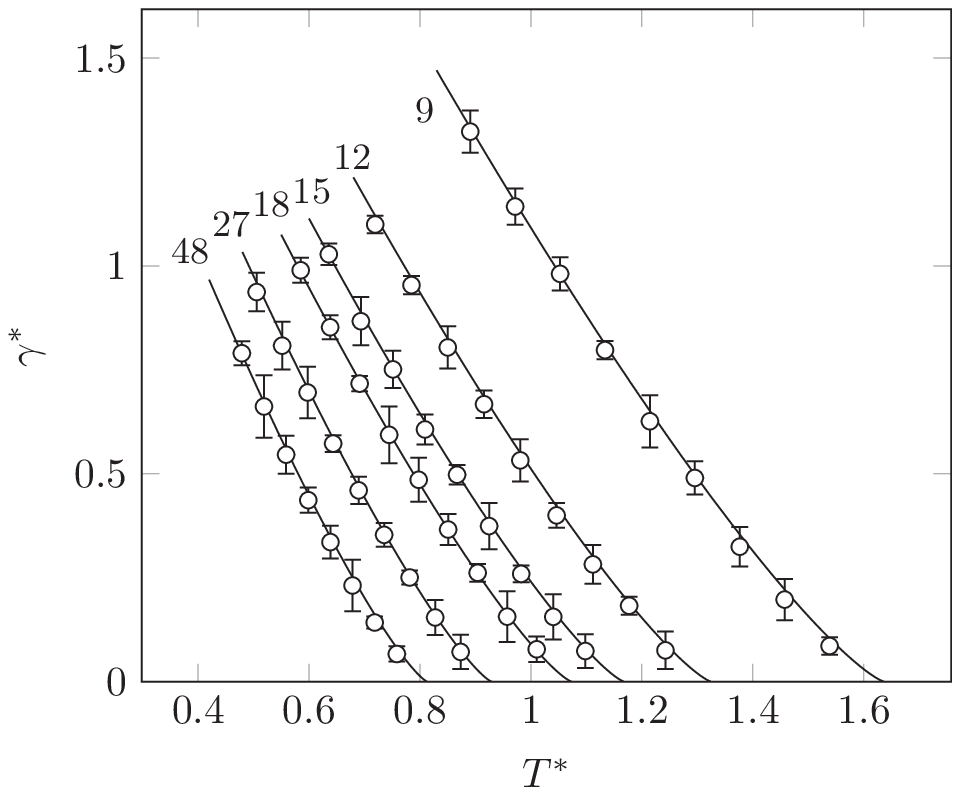}
\caption{Surface tension of the Mie fluid as a function of the temperature. The symbols are simulation results from the present work. The lines are correlations from Eq. (\ref{eq:Mie_Gamma}). The numbers are those for the exponent $n$ of the fluid.}  
\label{fig:Mie_gamma}
\end{figure}

Figs. \ref{fig:Mie_rhoT} - \ref{fig:Mie_gamma} show the results for the vapor pressure $p^\text{s*}$, the saturated liquid density $\rho^{'*}$, the saturated vapor density $\rho^{''*}$, the enthalpy of vaporization $\Delta h_{\rm V}^*$ and the surface tension $\gamma^{*}$ obtained for the Mie fluid with 14 different repulsive exponents. Further numerical details are given in Table \ref{tab:Mie_data}. The repulsive exponent $n$ has a strong influence of the VLE behavior. As $n$ increases, the critical temperature decreases and the slope of the coexistence curve changes. For higher values of $n$ the ratio of the triple point temperature and the critical temperature increases and therefore simulation were only performed for temperatures above 60 \% $T_\textrm{c}$ for $n \geq 36$.

\begin{longtable}{ll|lllll}
 \caption{Simulation results for the vapor pressure, the saturated densities, the enthalpy of vaporization and the surface tension of the Mie fluid from the present work. The numbers in parentheses indicate the uncertainties of the last decimal digits.} \\
\hline  \hline 
	      &   $T^*$ &  $p^\text{s*}$ &  $\rho^{'*}$ &  $\rho^{''*}$ & $\Delta h_{\rm V}^*$ & $\gamma^*$ \\
  \hline  \hline
$n = 9$ 	& 0.8907  & 0.0031(4) & 0.8147(6)  & 0.0035(4) & 7.62(1) & 1.31(5) \\ 
		& 0.9716  & 0.0062(6) & 0.7891(5)  & 0.0068(6) & 7.37(1) & 1.14(4) \\ 
		& 1.0526  & 0.0116(20)& 0.7559(8)  & 0.0118(19)& 7.08(1) & 0.98(4) \\
		& 1.1336  & 0.0191(12)& 0.7247(10) & 0.0189(12)& 6.73(3) & 0.80(2) \\
		& 1.2145  & 0.0301(13)& 0.6900(17) & 0.0291(11)& 6.31(2) & 0.63(6) \\ 
		& 1.2955  & 0.0458(35)& 0.6543(24) & 0.0448(34)& 5.77(3) & 0.49(4) \\ 
		& 1.3765  & 0.0645(37)& 0.6145(12) & 0.0639(41)& 5.15(5) & 0.32(5) \\
		& 1.4574  & 0.0894(28)& 0.5645(54) & 0.0924(40)& 4.37(8) & 0.20(5) \\
		& 1.5384  & 0.1193(21)& 0.5068(59) & 0.1344(97)& 3.33(21)& 0.09(2) \\
  \hline  
$n = 12$ 	& 0.7193  & 0.0019(3) & 0.8346(7)  & 0.0027(5) & 6.74(2) & 1.10(2) \\ 
		& 0.7847  & 0.0040(6) & 0.8061(5)  & 0.0053(7) & 6.50(2) & 0.95(2) \\ 
		& 0.8501  & 0.0078(9) & 0.7763(11) & 0.0098(14)& 6.24(2) & 0.80(5) \\
		& 0.9155  & 0.0136(11)& 0.7447(9)  & 0.0165(11)& 5.92(2) & 0.67(3) \\
		& 0.9809  & 0.0220(25)& 0.7107(22) & 0.0265(24)& 5.54(4) & 0.53(5) \\ 
		& 1.0463  & 0.0343(15)& 0.6745(39) & 0.0406(30)& 5.11(4) & 0.40(3) \\ 
		& 1.1117  & 0.0498(21)& 0.6325(38) & 0.0603(41)& 4.54(7) & 0.28(5) \\
		& 1.1771  & 0.0679(23)& 0.5831(39) & 0.0842(50)& 3.80(17)& 0.18(2) \\
		& 1.2425  & 0.0935(36)& 0.5225(93) & 0.129(13) & 2.67(51)& 0.08(5) \\
  \hline  
$n = 15$ 	& 0.6357  & 0.0013(3) & 0.8584(4)  & 0.0021(4) & 6.35(2) & 1.03(3) \\ 
		& 0.6935  & 0.0030(3) & 0.8296(5)  & 0.0044(5) & 6.15(1) & 0.87(6) \\ 
		& 0.7513  & 0.0059(5) & 0.7995(7)  & 0.0085(9) & 5.86(1) & 0.75(4) \\
		& 0.8091  & 0.0111(4) & 0.7675(5)  & 0.0149(10)& 5.59(3) & 0.61(4) \\
		& 0.8669  & 0.0177(14)& 0.7327(14) & 0.0237(17)& 5.23(3) & 0.50(2) \\ 
		& 0.9247  & 0.0278(28)& 0.6954(31) & 0.0367(45)& 4.85(4) & 0.37(6) \\ 
		& 0.9825  & 0.0414(19)& 0.6527(30) & 0.0555(30)& 4.35(4) & 0.26(2) \\
		& 1.0403  & 0.0598(10)& 0.6038(58) & 0.0837(29)& 3.70(5) & 0.16(5) \\
		& 1.0981  & 0.0817(40)& 0.5359(93) & 0.126(13) & 2.71(8) & 0.07(4) \\
  \hline  
$n = 18$ 	& 0.5850  & 0.0010(2) & 0.8801(4)  & 0.0017(3) & 6.16(2) & 0.99(3) \\ 
		& 0.6381  & 0.0024(7) & 0.8511(6)  & 0.0038(8) & 5.94(3) & 0.85(3) \\ 
		& 0.6913  & 0.0049(8) & 0.8201(5)  & 0.0075(10)& 5.68(3) & 0.72(2) \\
		& 0.7445  & 0.0091(13)& 0.7871(11) & 0.0135(15)& 5.39(3) & 0.59(7) \\
		& 0.7977  & 0.0157(16)& 0.7514(13) & 0.0227(20)& 5.05(3) & 0.49(5) \\ 
		& 0.8508  & 0.0249(20)& 0.7126(31) & 0.0357(33)& 4.65(3) & 0.37(4)\\ 
		& 0.9040  & 0.0364(26)& 0.6698(21) & 0.0528(51)& 4.18(7) & 0.26(2) \\
		& 0.9572  & 0.0496(40)& 0.6119(80) & 0.0818(39)& 3.57(9) & 0.16(6) \\
		& 1.0104  & 0.0755(44)& 0.5525(48) & 0.125(11) & 2.60(24)& 0.08(3) \\
  \hline  \hline
  \pagebreak
  \hline  \hline 
	      &   $T^*$ &  $p^\text{s*}$ &  $\rho^{'*}$ &  $\rho^{''*}$ & $\Delta h_{\rm V}^*$ & $\gamma^*$ \\
  \hline  \hline
$n = 21$ 	& 0.5503  & 0.0008(2) & 0.8984(2)  & 0.0015(3) & 6.03(1) & 0.96(9) \\ 
		& 0.6004  & 0.0019(1) & 0.8691(4)  & 0.0033(2) & 5.79(1) & 0.83(6) \\ 
		& 0.6504  & 0.0044(4) & 0.8376(12) & 0.0071(11)& 5.55(1) & 0.70(6) \\
		& 0.7004  & 0.0081(8) & 0.8039(9)  & 0.0126(16)& 5.25(2) & 0.59(6) \\
		& 0.7504  & 0.0143(14)& 0.7681(16) & 0.0215(30)& 4.94(5) & 0.48(2) \\ 
		& 0.8005  & 0.0228(8) & 0.7281(16) & 0.0346(8) & 4.54(1) & 0.35(5)\\ 
		& 0.8505  & 0.0339(22)& 0.6844(39) & 0.0515(44)& 4.10(6) & 0.26(3) \\
		& 0.9005  & 0.0507(21)& 0.6330(51) & 0.0804(47)& 3.45(6) & 0.16(3) \\
		& 0.9506  & 0.0701(16)& 0.565(13)  & 0.1213(72)& 2.62(16)& 0.07(2) \\
  \hline  
$n = 24$ 	& 0.5250  & 0.0007(2) & 0.9139(7)  & 0.0012(5) & 5.96(2) & 0.96(6) \\ 
		& 0.5727  & 0.0017(3) & 0.8842(6)  & 0.0031(6) & 5.70(4) & 0.82(7) \\ 
		& 0.6205  & 0.0038(7) & 0.8523(14) & 0.0065(8) & 5.45(4) & 0.71(5) \\
		& 0.6682  & 0.0074(5) & 0.8185(24) & 0.0122(10)& 5.17(3) & 0.58(5) \\
		& 0.7159  & 0.0128(10)& 0.7820(9)  & 0.0203(10)& 4.86(2) & 0.47(3) \\ 
		& 0.7636  & 0.0209(13)& 0.7416(15) & 0.0329(27)& 4.48(2) & 0.35(2)\\ 
		& 0.8114  & 0.0327(11)& 0.6965(31) & 0.0517(31)& 4.02(4) & 0.25(3) \\
		& 0.8591  & 0.0477(19)& 0.6444(60) & 0.0787(46)& 3.40(8) & 0.15(3) \\
		& 0.9068  & 0.0668(14)& 0.573(12)  & 0.121(11) & 2.57(17)& 0.07(3) \\
  \hline  
$n = 27$ 	& 0.5056  & 0.0006(2) & 0.9271(7)  & 0.0012(2) & 5.86(3) & 0.94(5) \\ 
		& 0.5516  & 0.0016(4) & 0.8972(6)  & 0.0030(4) & 5.63(3) & 0.81(6) \\ 
		& 0.5975  & 0.0036(3) & 0.8649(4)  & 0.0063(7) & 5.38(3) & 0.70(6) \\
		& 0.6435  & 0.0068(12)& 0.8304(12) & 0.0115(18)& 5.11(2) & 0.57(2) \\
		& 0.6895  & 0.0123(10)& 0.7931(15) & 0.0200(10)& 4.80(2) & 0.46(3) \\ 
		& 0.7354  & 0.0198(15)& 0.7516(15) & 0.0320(31)& 4.42(2) & 0.35(3) \\ 
		& 0.7814  & 0.0308(30)& 0.7069(15) & 0.0505(56)& 3.97(6) & 0.25(2) \\
		& 0.8274  & 0.0459(31)& 0.6530(48) & 0.0793(69)& 3.33(7) & 0.15(4) \\
		& 0.8733  & 0.0648(22)& 0.586(12)  & 0.1222(72)& 2.55(13)& 0.07(4) \\
  \hline  
$n = 30$ 	& 0.4902  & 0.0005(1) & 0.9383(4)  & 0.0010(1) & 5.80(2) & 0.96(4) \\ 
		& 0.5348  & 0.0015(5) & 0.9083(11) & 0.0028(8) & 5.57(3) & 0.83(3) \\ 
		& 0.5794  & 0.0032(4) & 0.8757(7)  & 0.0058(6) & 5.34(3) & 0.69(2) \\
		& 0.6239  & 0.0064(5) & 0.8405(12) & 0.0112(13)& 5.05(3) & 0.57(2) \\
		& 0.6685  & 0.0115(10)& 0.8027(9)  & 0.0194(17)& 4.74(2) & 0.47(4) \\ 
		& 0.7131  & 0.0188(15)& 0.7614(18) & 0.0311(30)& 4.38(2) & 0.35(3)\\ 
		& 0.7576  & 0.0295(14)& 0.7112(58) & 0.0510(21)& 3.87(5) & 0.24(3) \\
		& 0.8022  & 0.0448(6) & 0.6603(15) & 0.0789(19)& 3.31(3) & 0.16(3) \\
		& 0.8468  & 0.0623(16)& 0.5901(99) & 0.1194(22)& 2.58(8) & 0.07(2) \\
  \hline  \hline
  \pagebreak
  \hline  \hline 
	      &   $T^*$ &  $p^\text{s*}$ &  $\rho^{'*}$ &  $\rho^{''*}$ & $\Delta h_{\rm V}^*$ & $\gamma^*$ \\
  \hline  \hline
$n = 33$ 	& 0.4777  & 0.0005(2) & 0.9482(8)  & 0.0011(3) & 5.76(2) & 0.94(8) \\ 
		& 0.5212  & 0.0013(2) & 0.9179(12) & 0.0026(6) & 5.52(3) & 0.80(9) \\ 
		& 0.5646  & 0.0031(3) & 0.8848(9)  & 0.0057(5) & 5.29(3) & 0.69(3) \\
		& 0.6080  & 0.0061(6) & 0.8494(8)  & 0.0108(7) & 5.01(1) & 0.57(7) \\
		& 0.6515  & 0.0109(5) & 0.8105(6)  & 0.0187(6) & 4.70(2) & 0.46(4) \\ 
		& 0.6949  & 0.0180(6) & 0.7690(29) & 0.0309(13)& 4.33(3) & 0.35(2)\\ 
		& 0.7383  & 0.0293(15)& 0.7217(22) & 0.0509(18)& 3.86(3) & 0.24(2) \\
		& 0.7818  & 0.0429(13)& 0.6656(71) & 0.0776(50)& 3.28(7) & 0.15(5) \\
		& 0.8252  & 0.0613(20)& 0.5919(77) & 0.1206(63)& 2.50(9) & 0.08(3) \\
  \hline  
$n = 36$ 	& 0.5099  & 0.0013(3) & 0.9259(4)  & 0.0026(3) & 5.49(2) & 0.81(6) \\
		& 0.5524  & 0.0029(4) & 0.8927(8)  & 0.0056(6) & 5.23(2) & 0.69(5) \\ 
		& 0.5948  & 0.0058(3) & 0.8567(11) & 0.0104(7) & 4.98(1) & 0.56(3) \\
		& 0.6373  & 0.0102(11)& 0.8180(24) & 0.0183(15)& 4.65(3) & 0.44(4) \\
		& 0.6798  & 0.0176(14)& 0.7753(24) & 0.0308(38)& 4.29(3) & 0.35(6) \\ 
		& 0.7223  & 0.0282(17)& 0.7264(35) & 0.0499(27)& 3.83(4) & 0.25(3)\\ 
		& 0.7648  & 0.0424(29)& 0.6709(51) & 0.0786(75)& 3.24(7) & 0.14(3) \\
		& 0.8073  & 0.0614(38)& 0.6016(56) & 0.126(10) & 2.45(12)& 0.07(2) \\
  \hline  
$n = 39$ 	& 0.5003  & 0.0011(3) & 0.9332(15) & 0.0023(5) & 5.42(2) & 0.80(3) \\ 
		& 0.5420  & 0.0028(5) & 0.8996(6)  & 0.0053(5) & 5.22(2) & 0.68(5) \\ 
		& 0.5837  & 0.0055(7) & 0.8634(9)  & 0.0101(6) & 4.94(2) & 0.56(6) \\
		& 0.6254  & 0.0105(6) & 0.8240(13) & 0.0190(16)& 4.61(3) & 0.44(5) \\
		& 0.6671  & 0.0173(7) & 0.7807(22) & 0.0303(22)& 4.27(2) & 0.35(2) \\ 
		& 0.7088  & 0.0281(22)& 0.7325(32) & 0.0503(61)& 3.81(6) & 0.24(4)\\ 
		& 0.7505  & 0.0414(21)& 0.6765(41) & 0.0777(73)& 3.23(8) & 0.15(1) \\
		& 0.7922  & 0.0596(29)& 0.5999(78) & 0.124(11) & 2.42(12)& 0.06(3) \\
  \hline  
$n = 42$ 	& 0.4511  & 0.0011(2) & 0.9394(3)  & 0.0023(7) & 5.43(2) & 0.81(5) \\ 
		& 0.4922  & 0.0026(7) & 0.9058(10) & 0.0051(10)& 5.19(2) & 0.66(3) \\ 
		& 0.5332  & 0.0055(4) & 0.8689(13) & 0.0101(2) & 4.91(2) & 0.55(4) \\
		& 0.5742  & 0.0101(7) & 0.8289(13) & 0.0185(16)& 4.58(3) & 0.44(5) \\
		& 0.6152  & 0.0173(11)& 0.7851(16) & 0.0311(12)& 4.22(2) & 0.34(3) \\ 
		& 0.6972  & 0.0272(19)& 0.7374(28) & 0.0496(36)& 3.79(5) & 0.24(3)\\ 
		& 0.7382  & 0.0401(10)& 0.6791(52) & 0.0761(15)& 3.22(5) & 0.15(4) \\
		& 0.7793  & 0.0583(25)& 0.6028(61) & 0.1204(94)& 2.45(12)& 0.07(3) \\
  \hline  \hline
  \pagebreak
  \hline  \hline 
	      &   $T^*$ &  $p^\text{s*}$ &  $\rho^{'*}$ &  $\rho^{''*}$ & $\Delta h_{\rm V}^*$ & $\gamma^*$ \\
  \hline  \hline  
$n = 45$ 	& 0.4851  & 0.0011(3) & 0.9451(10) & 0.0023(4) & 5.41(2) & 0.79(7) \\ 
		& 0.5255  & 0.0026(4) & 0.9110(4)  & 0.0051(6) & 5.16(2) & 0.67(6) \\ 
		& 0.5659  & 0.0053(6) & 0.8745(10) & 0.0100(14)& 4.89(2) & 0.56(6) \\
		& 0.6064  & 0.0100(6) & 0.8342(23) & 0.0185(7) & 4.56(2) & 0.44(3) \\
		& 0.6468  & 0.0167(9) & 0.7901(38) & 0.0303(25)& 4.21(4) & 0.34(3) \\ 
		& 0.6872  & 0.0268(19)& 0.7411(37) & 0.0495(38)& 3.76(4) & 0.24(3)\\ 
		& 0.7276  & 0.0405(11)& 0.6824(28) & 0.0777(21)& 3.18(3) & 0.15(2) \\
		& 0.7681  & 0.0578(27)& 0.6054(83) & 0.1220(91)& 2.41(10)& 0.07(2) \\
  \hline
$n = 48$ 	& 0.4789  & 0.0011(3) & 0.9501(8)  & 0.0023(7) & 5.38(1) & 0.79(3) \\ 
		& 0.5188  & 0.0025(5) & 0.9157(12) & 0.0050(10)& 5.14(2) & 0.66(8) \\ 
		& 0.5587  & 0.0052(8) & 0.8786(14) & 0.0100(13)& 4.86(2) & 0.55(5) \\
		& 0.5987  & 0.0099(4) & 0.8383(14) & 0.0184(5) & 4.54(3) & 0.44(3) \\
		& 0.6386  & 0.0166(12)& 0.7935(12) & 0.0307(33)& 4.17(2) & 0.34(4) \\ 
		& 0.6785  & 0.0269(25)& 0.7441(29) & 0.0501(72)& 3.73(7) & 0.23(6)\\ 
		& 0.7184  & 0.0412(10)& 0.6838(42) & 0.0829(34)& 3.13(5) & 0.14(1) \\
		& 0.7583  & 0.0575(18)& 0.6099(73) & 0.1212(73)& 2.43(9) & 0.07(2) \\
  \hline
  \hline
%}
%\pagebreak 
\label{tab:Mie_data}
\end{longtable}

%All simulation results are correlated in order to be used for multi-criteria optimization of molecular models. 
For each value of the repulsive exponent $n$, the simulation results were correlated using the approach by Lotfi et al.~\cite{LVF92,KMWF95,SVHF01}. This approach is an extension of the density-temperature dependence given by Guggenheim \cite{Gugg45}, i.e. $\rho  \sim (T_{\rm c}-T)^{1/3}$. The saturated densities from simulation were described by
\begin{align}
 \rho'^{*} = \rho_{\rm c}^{*} + C_1 (T_{\rm c}^{*} - T^{*})^{1/3} + C_2' (T_{\rm c}^{*} - T^{*}) + C_3' (T_{\rm c}^{*} - T^{*})^{3/2}, 
 \label{eq:Mie_DensityLiquid}\\
 \rho''^{*} = \rho_{\rm c}^{*} - C_1 (T_{\rm c}^{*} - T^{*})^{1/3} + C_2'' (T_{\rm c}^{*} - T^{*}) + C_3'' (T_{\rm c}^{*} - T^{*})^{3/2}.
 \label{eq:Mie_DensityVapor}
\end{align}
The parameters $C_1$, $C_2'$, $C_2''$, $C_3'$, $C_3''$, as well as the critical densities and temperatures were fitted simultaneously to the simulation results and correlated as outlined below, cf. Eqs. (\ref{eq:Mie_Tc}) - (\ref{eq:Mie_X}). 

The vapor pressure was correlated using the approach by Lotfi et al.~\cite{LVF92,KMWF95,SVHF01}%, where the parameters are linear combinations...
\begin{equation}
 \ln p^{\rm s*} = c_1{}T^* - \frac{c_2}{T^*} - \frac{c_3}{T^{4*}}.
 \label{eq:Mie_VaporPressure}
\end{equation}
The enthalpy of vaporization was correlated using a similar approach as for the saturated liquid density. The first parameter was omitted, so that the enthalpy of vaporization decreases towards zero at the critical point.
\begin{equation}
 \Delta h_{\rm V}^{*} = d_1 (T_{\rm c}^{*} - T^{*})^{1/3} + d_2 (T_{\rm c}^{*} - T^{*}) + d_3 (T_{\rm c}^{*} - T^{*})^{3/2}
 \label{eq:Mie_Enthalpy}
\end{equation}
Following the principle of corresponding states \cite{Gugg45,MHM14,G10}, the surface tension was correlated here using a critical scaling expression
\begin{equation}
 \gamma^* = A \left(1-\frac{T^*}{T_{\rm c}^*}\right)^B.
 \label{eq:Mie_Gamma}
\end{equation}
The individual parameters are correlated as global functions of the repulsive exponent $n$. It turns out that for correlating the critical temperature and density, the simple forms presented in Eqs. (\ref{eq:Mie_Tc}) and (\ref{eq:Mie_rhoc}) give good results:
\begin{equation}
 T_{\rm c} = a + b/n+c/n^3,
 \label{eq:Mie_Tc}
\end{equation}
\begin{equation}
 \rho_{\rm c} = d + e \log(n).
 \label{eq:Mie_rhoc}
\end{equation}
The numbers for $a, b, c, d,$ and $e$ are given in Table \ref{tab:Mie_fitCrit}.

\begin{table}[htb]
\centering
%\resizebox{0.75 \textwidth}{!}{
 \caption{Parameters for critical properties from Eqs. (\ref{eq:Mie_Tc}) and (\ref{eq:Mie_rhoc}) fit to the present simulation results.}
 \begin{tabular}{lll | lll} \\
  \hline  \hline
  $T_{\rm c}$	&		&					&	$\rho_{\rm c}$ \\
  \hline
  $a$	& 1		& 0.65978 				&	$d$	& 1		&	0.25325  \\
  $b$	& $1/n$		& 0.69171 $ \cdot \, 10^{1}$		&	$e$	& $\log(n)$	&	0.58291 $ \cdot \, 10^{-1}$ \\
  $c$	& $1/n^3$	& 0.14260 $ \cdot \, 10^{3}$		&		& 		&	 \\
  \hline  \hline
 \end{tabular}
 \label{tab:Mie_fitCrit}
\end{table}

The individual parameters of the correlations in Eqs. (\ref{eq:Mie_DensityLiquid}) - (\ref{eq:Mie_Gamma}) can be described by similar functional forms dependent on the repulsive exponent $n$, cf. Eq. \ref{eq:Mie_X}:
\begin{equation}
 X = \alpha_X + \beta_X \, n + \eta_X \, / \, n + \delta_X \, \log(n),
 \label{eq:Mie_X}
\end{equation}
where $X$ is any of the correlation parameters introduced above. The individual parameters $\alpha_X$, $\beta_X$, $\eta_X$, $\delta_X$ for Eq. (\ref{eq:Mie_X}) are given in Table \ref{tab:Mie_fit}.

\begin{table}[htb]
\centering
 \caption{Parameters for the correlations from Eqs. (\ref{eq:Mie_DensityLiquid}) - (\ref{eq:Mie_Gamma}), adjusted to the present simulation results.}
 \begin{tabular}{ll|l|l|l} \\
  \hline  \hline
  Saturated liquid density&		& $C_1$					&	$C_2'$				& $C_3'$ \\
  \hline
  $\alpha$	& 1		&$-0.13398  \cdot \, 10^{1}$		&	$\phantom{-}0.15206  \cdot \, 10^{1}$	&$\phantom{-}0.16785  \cdot \, 10^{-2}$\\
  $\beta$	& $n$		&$-0.13412  \cdot \, 10^{-1}$		&	$\phantom{-}0.14540 \cdot \, 10^{-1}$	&  \\
  $\eta$	& $1/n$		&$\phantom{-} 0.36607 \cdot \, 10^{1}$	&	$-0.46863  \cdot \, 10^{1}$		& $\phantom{-}0.53675$  \\
  $\delta$	& $\log(n)$	&$\phantom{-} 0.15206 \cdot \, 10^{1}$	&	$-0.91705 $				& $-0.11174$ \\
  \hline  
  Saturated vapor density&		& $C_2''$				&	$C_3''$					& \\
  \hline
  $\alpha$	& 1		&$-0.32782 \cdot \, 10^{1}$		&	$-0.72676 \cdot \, 10^{-2}$	& \\
  $\beta$	& $n$		&$-0.24242 \cdot \, 10^{-1}$		&						& \\
  $\eta$	& $1/n$		&$\phantom{-} 0.83828 \cdot \, 10^{1}$	&						& \\
  $\delta$	& $\log(n)$	&$\phantom{-} 0.26264 \cdot \, 10^{1}$	&	$\phantom{-}0.17001$ 			& \\
  \hline  
  Vapor pressure&		& $c_1$					&	$c_2$					& $c_3$ \\
  \hline
  $\alpha$	& 1		&$\phantom{-} 0.65036 \cdot \, 10^{1}$	&	$\phantom{-}0.22655 \cdot \, 10^{1}$	&$\phantom{-} 0.79367  \cdot \, 10^{-1}$ \\
  $\beta$	& $n$		&$\phantom{-} 0.60373 \cdot \, 10^{-1}$	&	$\phantom{-}0.12654 \cdot \, 10^{-1}$	&$-0.10051  \cdot \, 10^{-2}$ \\
  $\eta$	& $1/n$		&$-0.20838 \cdot \, 10^{2}$		&	$\phantom{-}0.30984 \cdot \, 10^{2}$	&$\phantom{-} 0.31402$ 	 \\
  $\delta$	& $\log(n)$	&$-0.38117 \cdot \, 10^{1}$		&	$-0.82858 \cdot \, 10^{-3}$		& \\
  \hline  
  Enthalpy of vaporization&	& $d_1$					&	$d_2$					& $d_3$ \\
  \hline
  $\alpha$	& 1		&$-0.20824 \cdot \, 10^{1}$		&	$\phantom{-}0.17862 \cdot \, 10^{2}$	&$-0.46053$ \\
  $\beta$	& $n$		&$-0.76099 \cdot \, 10^{-1}$		&	$\phantom{-}0.11423$ 			&$\phantom{-} 0.11186$ \\
  $\eta$	& $1/n$		& 					&						& 	 \\
  $\delta$	& $\log(n)$	&$\phantom{-} 0.60306 \cdot \, 10^{1}$	&	$-0.98931 \cdot \, 10^{1}$		&$-0.53200 \cdot \, 10^{1}$\\
  \hline  
  Surface tension&		& $A$					&	$B$					&  \\
  \hline
  $\alpha$	& 1		&$-0.99270 \cdot \, 10^{1}$		&	$\phantom{-}0.12572 \cdot \, 10^{1}$	&  \\
  $\beta$	& $n$		&$-0.71259 \cdot \, 10^{-1}$		&						&  \\
  $\eta$	& $1/n$		&$\phantom{-} 0.52405 \cdot \, 10^{2}$	&						&  \\
  $\delta$	& $\log(n)$	&$\phantom{-} 0.87240 \cdot \, 10^{1}$	&						&  \\  
  \hline \hline
 \end{tabular}
 \label{tab:Mie_fit}
\end{table}

%Figs. \ref{fig:Mie_rhoT} to \ref{fig:Mie_gamma} show the simulation results for thermodynamic properties of 6 Mie fluids. 
The correlations represent the simulation results within the statistical uncertainties in most cases, cf. Figs. \ref{fig:Mie_rhoT} - \ref{fig:Mie_gamma}.

\begin{figure}[htb]
\centering
\resizebox{0.62 \textwidth}{!}{
 \includegraphics{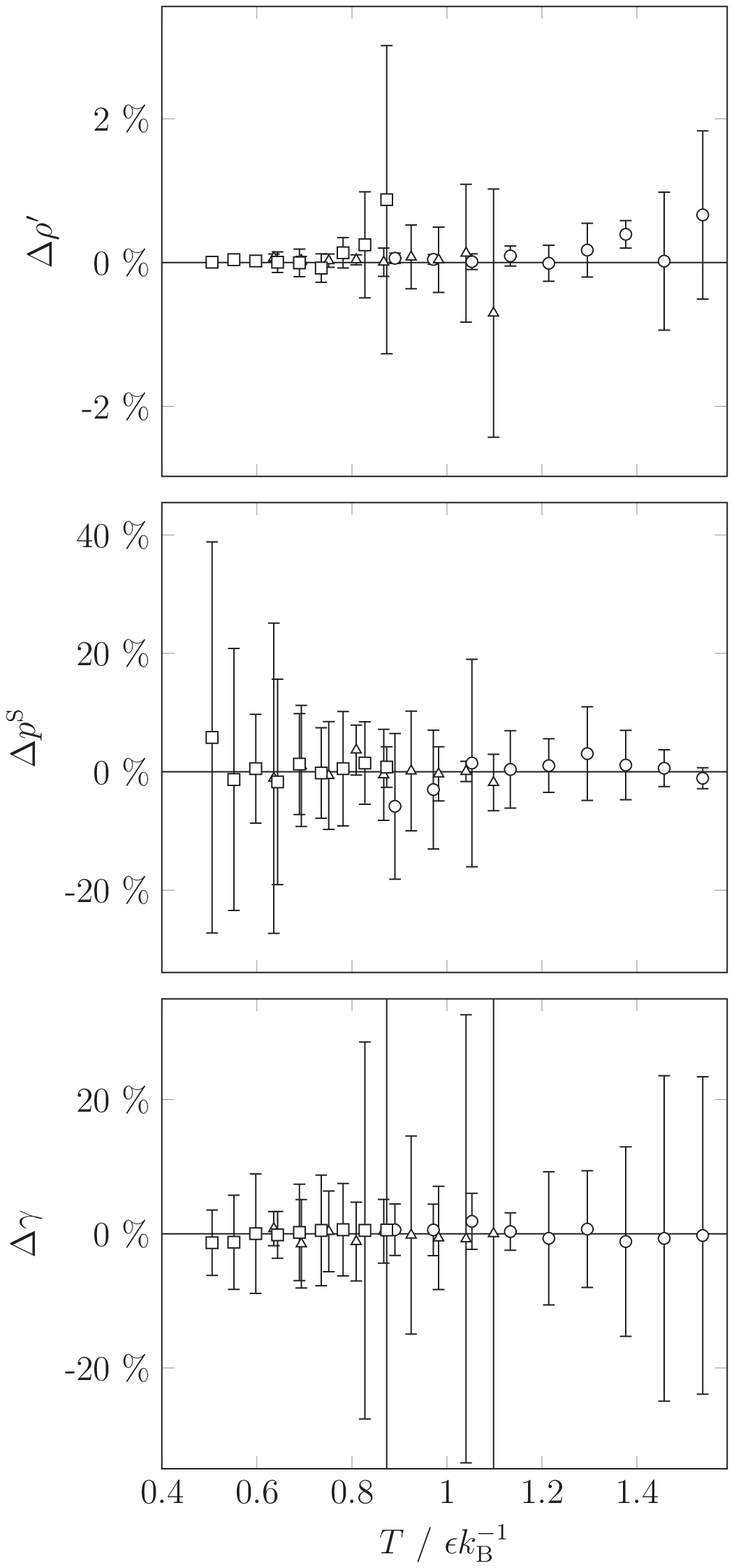}
 }
\caption{Relative deviation $\Delta X = (X_{\rm corr} - X_{\rm sim}) / X_{\rm sim}$ of simulation results of the Mie fluid from the correlations given by Eq.\ (\ref{eq:Mie_DensityLiquid}) (top), Eq.\ (\ref{eq:Mie_VaporPressure}) (center) and Eq.\ (\ref{eq:Mie_Gamma}) (bottom), for different repulsive exponents: 9 (\Circle), 15 ($\triangle$), 27 ($\square$).}  
\label{fig:Mie_dev}
\end{figure}

Fig.\ \ref{fig:Mie_dev} shows the relative deviation of the simulation results from the correlations for the Mie fluid with $n = 9$, $n = 15$ and $n = 27$. The relative mean deviations of the correlations from simulation data is calculated by
\begin{equation}
 \delta X = \sqrt{\frac{1}{N} \frac{1}{M} \sum_{i}^N \sum_{j}^M \left(\frac{(X_{i,\rm corr}(T_j) - X_{i,\rm sim}(T_j)}{ X_{i,\rm sim}(T_j)}\right)^2},
\end{equation}
where $i$ represents a counter for the different exponents of the Mie fluid and $j$ the different temperatures. The numerical values for the relative mean deviations are 0.19 \% for the saturated liquid density, 9.8 \% for the saturated vapor density, 1.6 \% for the vapor pressure, 0.84 \% for the enthalpy of vaporization and 1.9 \% for the surface tension. The correlation for the saturated vapor density does not capture the limiting case of the ideal gas and should therefore not be used for temperatures below 0.7 $T_{\rm c}$ \cite{SVHF01}.

\begin{figure}[htb]
\centering
\resizebox{0.55 \textwidth}{!}{
 \includegraphics{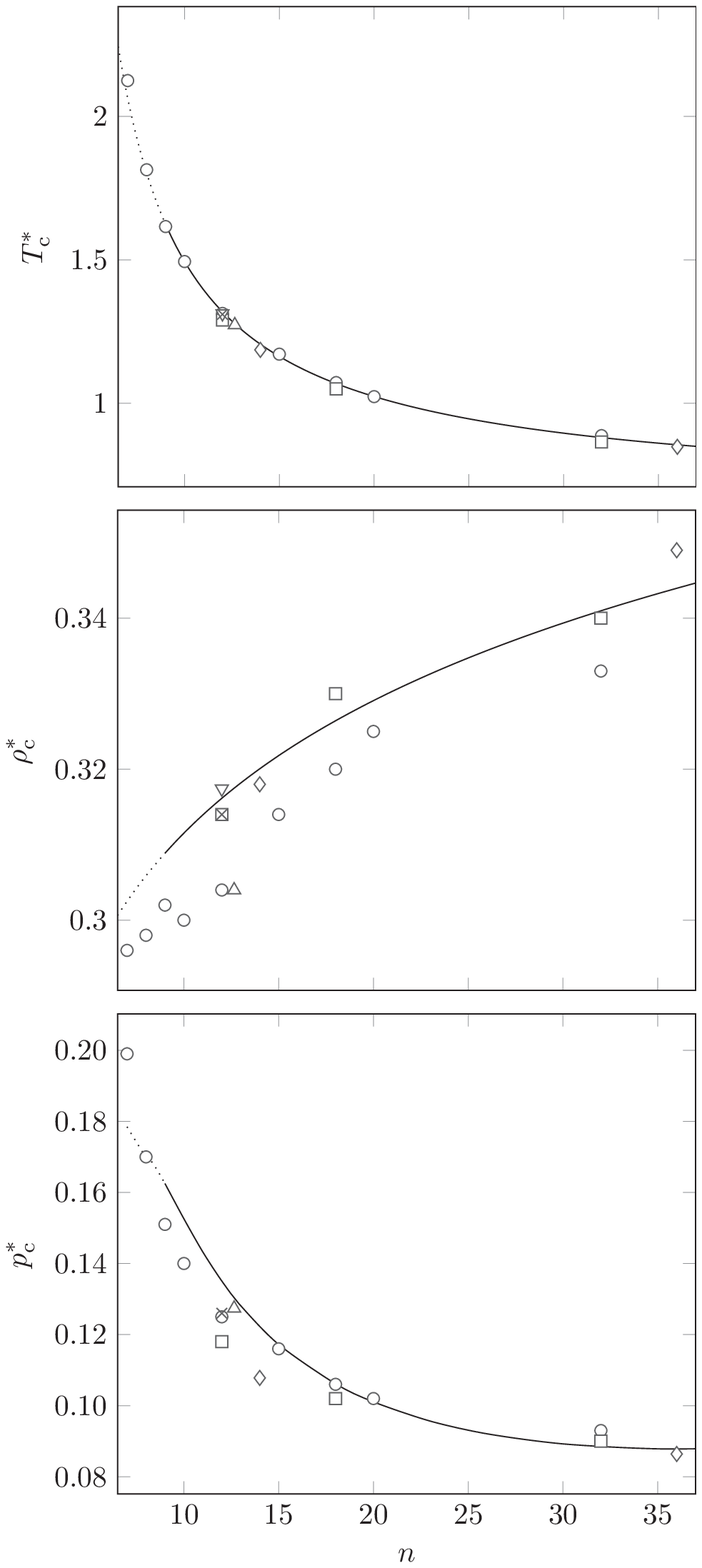}
 }
\caption{Critical properties of the Mie fluid as a function of the repulsive exponent: critical temperature (top), critical density (center) and critical pressure (bottom). The solid lines are the correlations given in Eq. (\ref{eq:Mie_Tc}), (\ref{eq:Mie_rhoc})  and (\ref{eq:Mie_VaporPressure}), and the symbols are results from the work of: Okumura and Yonezawa \cite{OY00} (\Circle), Orea et al. \cite{ORD08} ($\square$), Lafitte et al. \cite{LAAGAMJ13} ($\triangle$), Potoff and Bernard-Brunel \cite{PB09} ($\diamond$), Lotfi et al. \cite{LVF92} ($\times$), P\'{e}rez-Pellitero et al. \cite{PPUOM06} ($\triangledown$). The present correlation is continued as a dotted line for $n < 9$.}  
\label{fig:Mie_Tc}
\end{figure}

Fig. \ref{fig:Mie_Tc} shows a comparison of the present critical data with results from the literature. As the repulsive exponent $n$ increases, the critical temperature decreases. The correlation given by Eq. (\ref{eq:Mie_Tc}) is in very good agreement with the literature data, even for $n < 9$, i.e. in a range to which it was not adjusted. The critical density increases only very slightly with the repulsive exponent and this trend is well reproduced by the correlation and confirmed by literature data. The results by Okumura and Yonezawa \cite{OY00} for the critical density are smaller than the present values, but the correlation agrees well with literature data on the Lennard-Jones fluid \cite{LVF92,PPUOM06}. The critical pressure was determined with a combination of Eqs. (\ref{eq:Mie_VaporPressure}) and (\ref{eq:Mie_Tc}). As the repulsive exponent $n$ increases, the critical pressure decreases, which is well predicted by the correlation and confirmed by literature data.

\begin{figure}[htb]
\centering
 \includegraphics{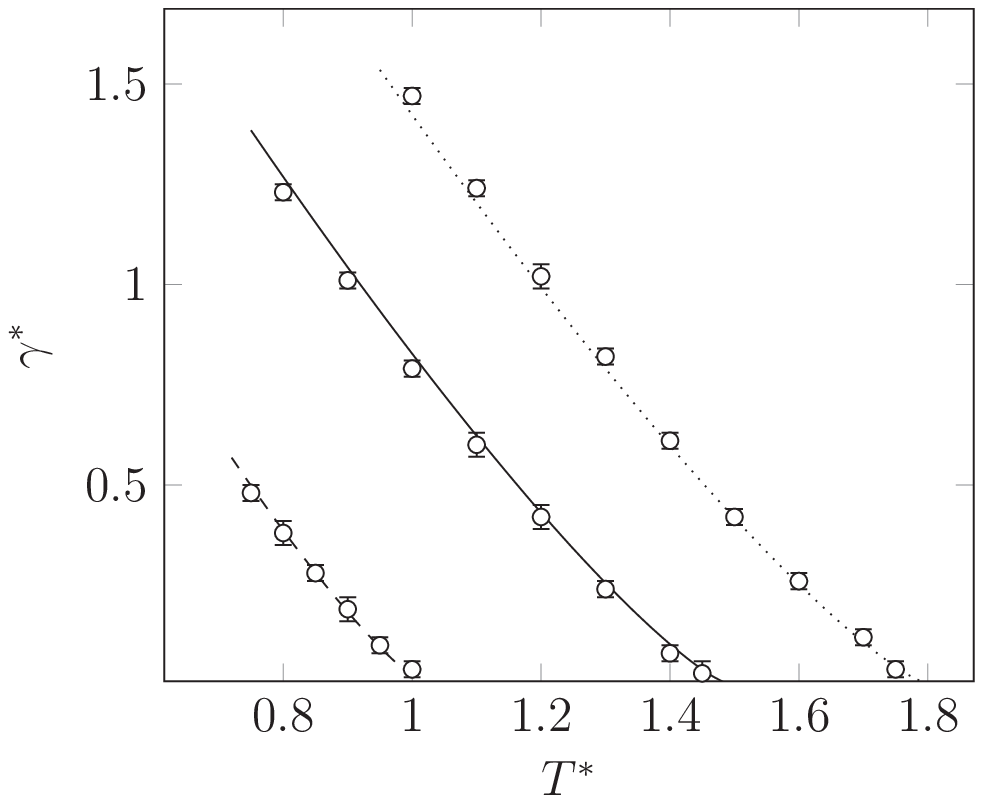}
\caption{Surface tension of the Mie fluid as a function of  the temperature. The symbols are simulation results from Galliero et al. \cite{GPMMLB09} and the lines are correlations from Eq. (\ref{eq:Mie_Gamma}): $n = 8$ ($\cdots$), $n = 10$ (\solidrule) and $n = 20$ (\dashedrule).}  
\label{fig:Mie_gammaComp}
\end{figure}

\begin{figure}[htb]
\centering
 \includegraphics{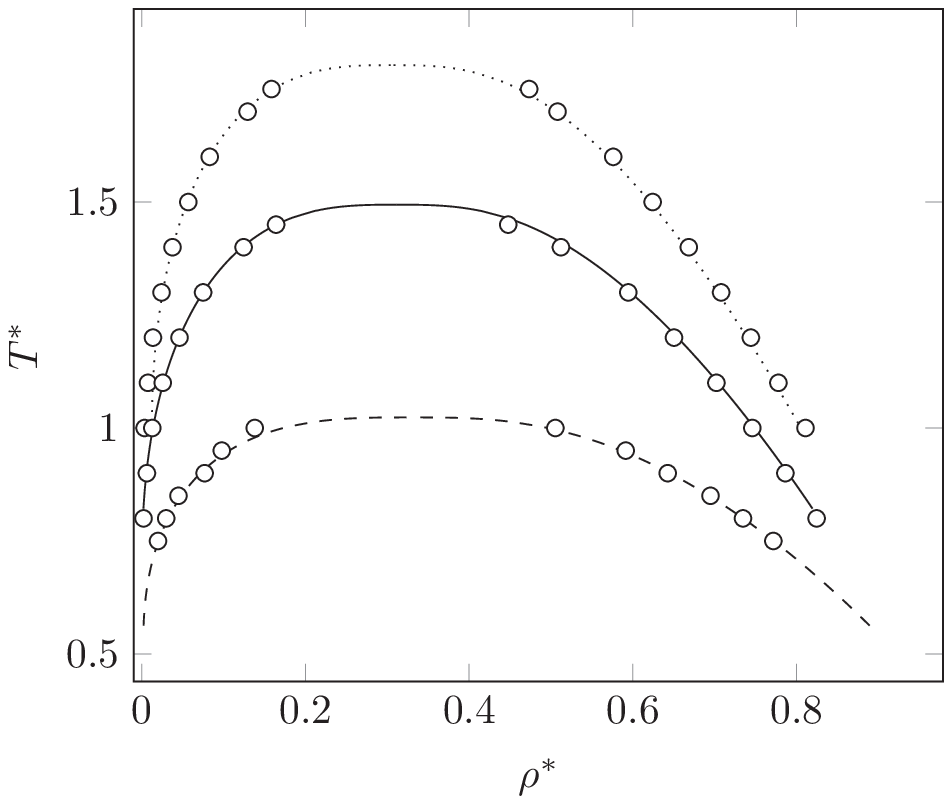}
\caption{Saturated densities of the Mie fluid. The symbols are simulation results from Galliero et al. \cite{GPMMLB09} and the lines are correlations from Eq. (\ref{eq:Mie_Gamma}): $n = 8$ ($\cdots$), $n = 10$ (\solidrule) and $n = 20$ (\dashedrule).}  
\label{fig:Mie_densComp}
\end{figure}

Fig. \ref{fig:Mie_gammaComp} shows the surface tension of the Mie fluid over the temperature and Fig. \ref{fig:Mie_densComp} shows the saturated densities of the Mie fluid. The simulation results from the present work are in very good agreement with the simulation data of Galliero et al. \cite{GPMMLB09}, even though the correlation is extrapolated  to $n < 9$ here as well. %Galliero et al. \cite{GPMMLB09} used a cutoff radius of $r_{\rm c} = 10$ $\sigma$, which proves that the present LRC approach is an efficient way to treat the long-range interactions.

\section{Application to carbon dioxide and comparison with other potentials}

The correlations introduced above can be used to adjust model parameters to experimental VLE data. In the present work, the Mie fluid (with three parameters $\sigma$, $\epsilon$ and $n$) is compared to the standard 12-6 Lennard-Jones potential (with two parameters $\sigma$ and $\epsilon$) and the two-center Lennard-Jones plus point quadrupole (2CLJQ) model (with four parameters $\sigma$, $\epsilon$, the elongation $L$ and the quadrupole moment $Q$). Carbon dioxide is used as a test case. There are several 2CLJQ models \cite{VSH01,WSKKHH15,MF94,MSM81,MSB76} and three site models with superimposed electrostatics \cite{HY95,ZD05,MEVH10,JMEP16} for the description of CO$_2$ and one model based on a single-site four-parameter Mie potential \cite{ALGAJM11}.

Following the approach introduced by St\"obener et al. \cite{SKRMKH14}, multi-criteria optimization based on Pareto sets is used in the present study. The three objective functions $\delta O$ are considered here, representing relative mean deviations in the saturated liquid density, the vapor pressure and the surface tension,
\begin{equation}
 \delta O = \sqrt{\frac{1}{K}\sum_{j=1}^K \left(\frac{O^{\rm exp}(T_j)-O^{\rm sim}(T_j,\sigma,\epsilon,n,L,Q)}{O^{\rm exp}(T_j)}\right)^2},
\end{equation}
where the $O^{\rm exp}$ are calculated by DIPPR correlations to experimental data, and the $O^{\rm sim}$ are obtained from Eqs. (\ref{eq:Mie_DensityLiquid}), (\ref{eq:Mie_VaporPressure}) and (\ref{eq:Mie_Gamma}). The thermodynamic properties were evaluated at 15 temperatures $T_j$ from the triple point temperature up to 95 \% of the critical temperature of carbon dioxide ($T_{\rm c}$ = 304.13 K \cite{SW96}) in equal steps.

The DIPPR correlations as well as the correlations to the simulation data are subject to errors. As discussed above, the relative mean deviations of the correlations to simulation data are 0.2 \%, 1.6 \% and 1.9 \% for saturated liquid density, the vapor pressure and the surface tension, respectively. The corresponding relative mean deviations for the DIPPR correlations are 0.2 \%, 1 \% and 4 \%, respectively \cite{DIPPR}. 
%\textbf{The relative deviations of the correlations to simulation data for the 2CLJQ model fluid are reported to be below 0.1 \% for low temperatures and up to 3 \% in the critical region, the vapor pressures deviate up to 20 \% at low temperatures and below 2 \% at higher temperatures \cite{SVHF01}(\"Uberarbeiten!)}. The surface tension correlation shows a very similar performance compared to the correlation for the Mie model fluid, resulting in a relative mean deviation of 1.9 \% \cite{WHH15a}.

The Pareto set for the multicriteria optimization problem described above was determined in different ways depending on the potential: For the LJ and the Mie model, a brute force sampling of the parameter space was performed \cite{WSKKHH15}, while the 2CLJQ Pareto set was determined by a combination of sandwiching and hyperboxing, for details see St\"obener et al. \cite{SKHKH15}.

For the brute force sampling of the parameters $\sigma$, $\epsilon$ and $n$, a sample grid consisting of 200 $\times$ 200 $\times$ 50 points was used. For the Lennard-Jones potential, the repulsive exponent was fixed and the grid size for $\sigma$ and $\epsilon$ was 200 $\times$ 200.

\begin{figure}[htb]
\centering
 \includegraphics{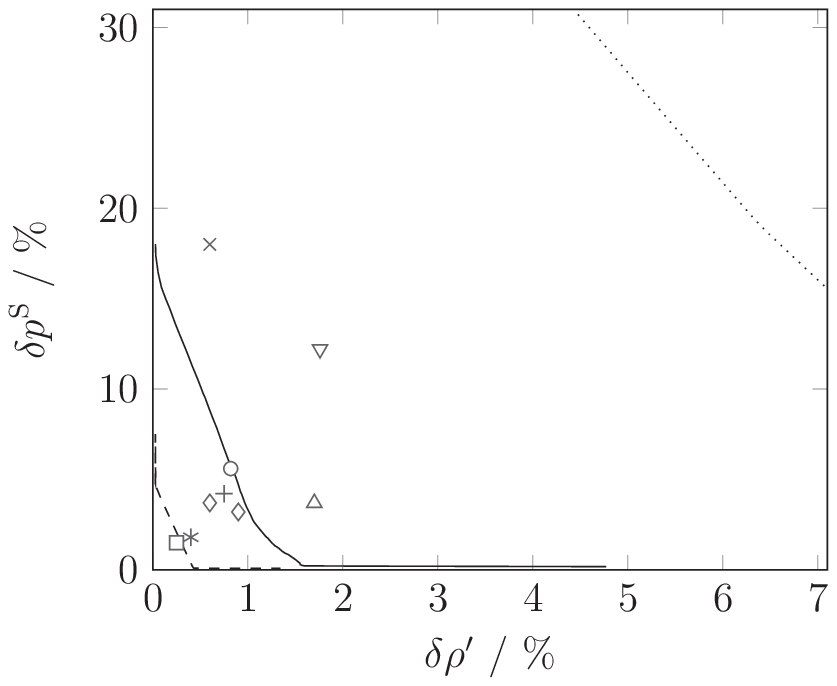}
\caption{Pareto set for CO$_2$. The dotted line corresponds to the Lennard-Jones fluid, the solid line to the three-parameter Mie fluid, and dashed line to the 2CLJQ fluid. The symbols represent molecular models: ($\triangledown$) Harris and Yung (3CLJ + 3 partial charges) \cite{HY95}, ($\times$) Zhang and Duan (3CLJ + 3 partial charges) \cite{ZD05,MVH08}, ($\diamond$) Jiang et al. \cite{JMEP16} (3 Buckingham sites + 3 Gaussian charges), ($\ast$) Merker et al. \cite{MEVH10} (3CLJQ), ($\square$) Vrabec et al. (2CLJQ) \cite{VSH01}, ($+$) M\"oller and Fischer (2CLJQ) \cite{MF94}, ($\triangle$) Avenda\~{n}o et al. (four-parameter Mie) \cite{ALGAJM11,ALAGMJ13} and (\Circle) the present model with $\sigma$ = 3.768 $\textup{\AA}$, $\epsilon / k_{\rm B}$  = 366 K and $n$ = 39.}  
\label{fig:Mie_rhopCO2}
\end{figure}

Fig. \ref{fig:Mie_rhopCO2} shows the Pareto set determined with respect to two objective functions: the deviation in the saturated liquid density $\delta \rho'$ and the vapor pressure $\delta p^{\rm S}$. All model parameterizations which are Pareto-optimal for the two-criteria optimization remain Pareto-optimal if the third optimization criterion, i.e. the surface tension, is taken into account. The three different lines correspond to the three different molecular model types. The Lennard-Jones model yields very large errors and is obviously not suited for describing the studied properties of CO$_2$. It is therefore not discussed further.

The Pareto of the two models as depicted in Fig. \ref{fig:Mie_rhopCO2} reveal the typical features: they include different regions: extreme compromises and the so-called Pareto knee. As an extreme compromise, the deviation in saturated liquid density can be below 0.03 \%, however only at the expense of deviations in the vapor pressure of 15 \% (2CLJQ) or 18 \% (Mie). The deviation in the vapor pressure can be below 0.4 \% if deviations in the saturated liquid density of 4 \% (2CLJQ) or 5 \% (Mie) are accepted. %Both scenarios represent extreme compromises where one property is represented within the statistical uncertainties of the experimental data and the other property shows much larger deviations.
The most attractive part of the Pareto set is usually in between those extreme cases, i.e. in the Pareto knee. The repulsive exponent $n$ varies between 24 and 39, which compares favorably with Avenda\~{n}o et al. \cite{ALGAJM11}, who used a repulsive exponent of $n$ = 23, but a different dispersive exponent, and Maurer \cite{M78}, who used exponents of $n$ = 30 and $n$ = 33 for a perturbation theory. Ramrattan et al. concluded that a repulsive exponent $n=31$ should be used for an attractive exponent of $m=6$ \cite{RAMG15}. %The slope of the Pareto set changes dramatically in the ''Pareto knee``, which makes it usually hard to optimize molecular models. 
%Fig. \ref{fig:Mie_rhopCO2} gives an overview how well each of the considered model classes can reproduce the experimental data of CO$_2$. The Pareto sets of the different molecular model classes do not intersect, which makes it easy to determine which class represents the experimental data in the best way. The single-site Lennard-Jones model does not reasonably reproduce the experimental data. Varying the repulsive exponent enables a representation of the experimental data with a good agreement, i.e. deviations in the saturated liquid density below 1 \% and the vapor pressure below 6 \% simultaneously. 
Comparing the Pareto set of the Mie and the 2CLJQ model shows that the latter enables a better description of the studied data of CO$_2$. Fig. \ref{fig:Mie_rhopCO2} also includes results from some literature models of CO$_2$.
 
%The model by Avenda\~{n}o et al. \cite{ALGAJM11} is not part of the two dimensional Pareto set, but it is very close. It should be noted here that 
Avenda\~{n}o et al. \cite{ALGAJM11} used a 4 parameter Mie model in which the dispersive exponent of $n$ = 6.66, which was adjusted. %(i.e. four rather than three free model parameters were used),\cite{ALGAJM11} instead of the exponent $n$ = 6 in the present work. 
The optimization of the model by Avenda\~{n}o et al. \cite{ALGAJM11} was performed not only with a focus on the vapor pressure and the saturated liquid density, but also on the surface tension and transport properties. 
%The 2CLJQ model class is able to reduce the deviations in the objective functions: deviations of 0.2 \% in the saturated liquid density and 1.5 \% in the vapor pressure are reported by Vrabec et al. \cite{VSH01}.
The model by Vrabec et al. \cite{VSH01} is a reparameterization of the 2CLJQ model by M\"oller and Fischer \cite{MF94}. %, which shows the potential of the 2CLJQ model for describing the VLE of carbon dioxide. 
More complex models in the literature are based on three Lennard-Jones sites and three partial charges, e.g. \cite{HY95,ZD05}. These models do not necessarily represent the VLE with a higher accuracy, but instead show higher deviations than the single-site Mie model. Very recently Jiang et al. \cite{JMEP16} developed two Buckingham potential models for CO$_2$ based on Gaussian charges. These models perform very well for transport properties and homogeneous state points, whereas no improvement for description of the vapor-liquid equilibrium is found \cite{JMEP16}.  %The complexity of the molecular models is higher with two Lennard-Jones sites and a superimposed point quadrupole. 

\begin{figure}[htb]
\centering
 \includegraphics{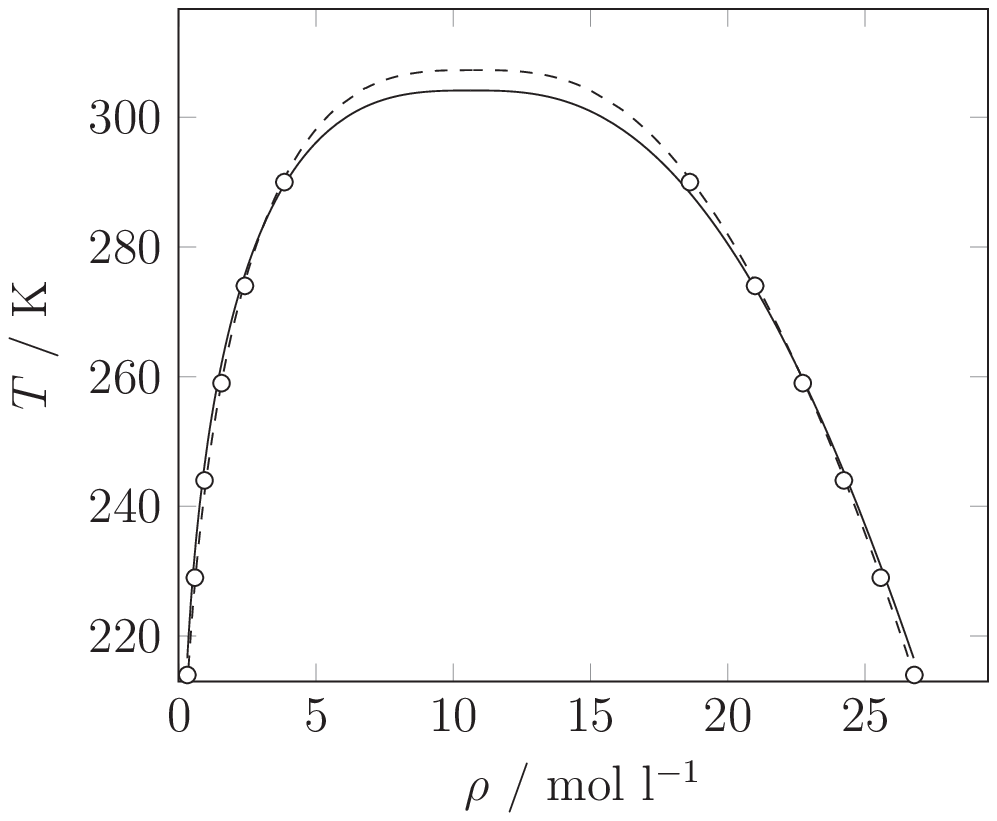}
\caption{Saturated densities of CO$_2$. The symbols are the present simulations results, the dashed line represents Eqs. (\ref{eq:Mie_DensityLiquid}) and (\ref{eq:Mie_DensityVapor}), and the solid line represents correlations to experimental data \cite{SW96}. The simulation uncertainties are smaller than the symbol size in all cases.}  
\label{fig:Mie_rhoTCO2}
\end{figure}

\begin{figure}[htb]
\centering
 \includegraphics{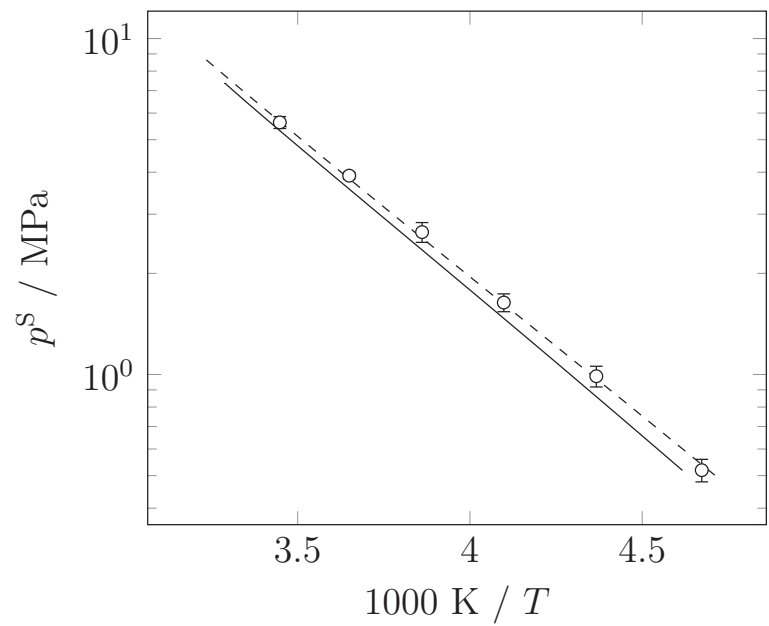}
\caption{Vapor pressure curve of CO$_2$. The symbols are the present simulations results, the dashed line represents Eq.\ (\ref{eq:Mie_VaporPressure}) and the solid line represents a correlation to experimental data \cite{SW96}.}  
\label{fig:Mie_pTCO2}
\end{figure}

\begin{figure}[htb]
\centering
 \includegraphics{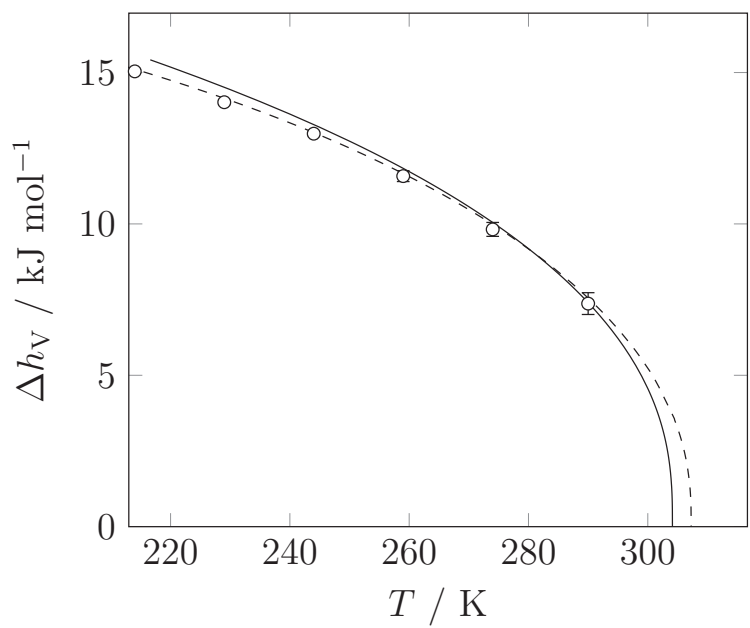}
\caption{Enthalpy of vaporization of CO$_2$ as a function of  the temperature. The symbols are the present simulations results, the dashed line represents Eq. (\ref{eq:Mie_Enthalpy}) and the solid line represents a correlation to experimental data \cite{SW96}.}  
\label{fig:Mie_deltahVCO2}
\end{figure}

\begin{figure}[htb]
\centering
 \includegraphics{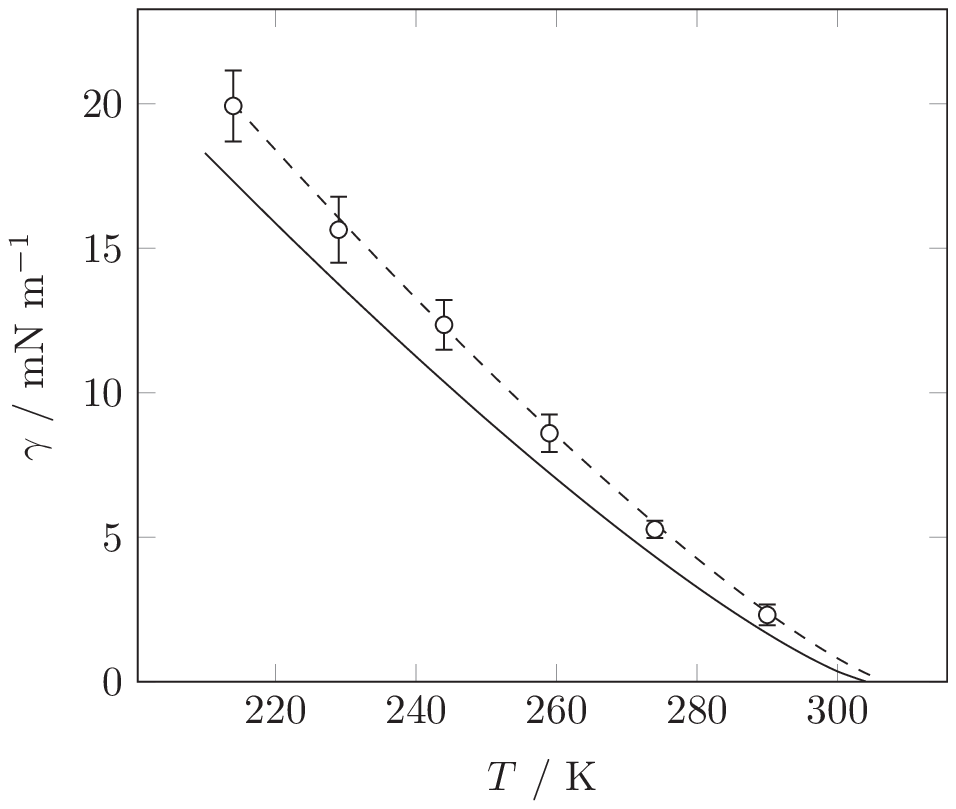}
\caption{Surface tension of CO$_2$ as a function of the temperature. The symbols are the present simulation results, the dashed line represents Eq.\ (\ref{eq:Mie_Gamma}) and the solid line represents a correlation to experimental data \cite{DIPPR}.}  
\label{fig:Mie_gammaCO2}
\end{figure}

As one possible compromise between the two objectives of minimizing $\delta p^{\rm S}$ and $\delta \rho'$, a model parameterization is selected here from the Pareto set which reaches an overall agreement of $\delta \rho' = 0.8$ \%, $\delta p^{\rm S} = 5.6$ \%, $\delta \Delta h_{\rm V} = 2.1$ \% and $\delta \gamma = 23.3$ \%. The corresponding model parameters are $\sigma$ = 3.768 $\textup{\AA}$, $\epsilon$ / $k_{\rm B}$  = 366 K and $n$ = 39.  The representation of the discussed thermodynamics properties of the present Mie model is shown in Figs. \ref{fig:Mie_rhoTCO2} to \ref{fig:Mie_gammaCO2}. The saturated liquid density and the vapor pressure were used for the parameterization of the Mie fluid and therefore both properties show a good agreement with experimental data. The enthalpy of vaporization was not used in the parameterization, but the predictions match the experimental data well. 
The surface tensions are also predictions. There are deviations from experimental data larger than 20 \%, which is typical for molecular models which are adjusted in a similar manner \cite{WSKKHH15,SKHKH15,EV15,GMT14,NWLM12,LALMJ15,ZLCMSA13,ATC95}.

\begin{figure}[htb]
\centering
\resizebox{1.0 \textwidth}{!}{
 \includegraphics{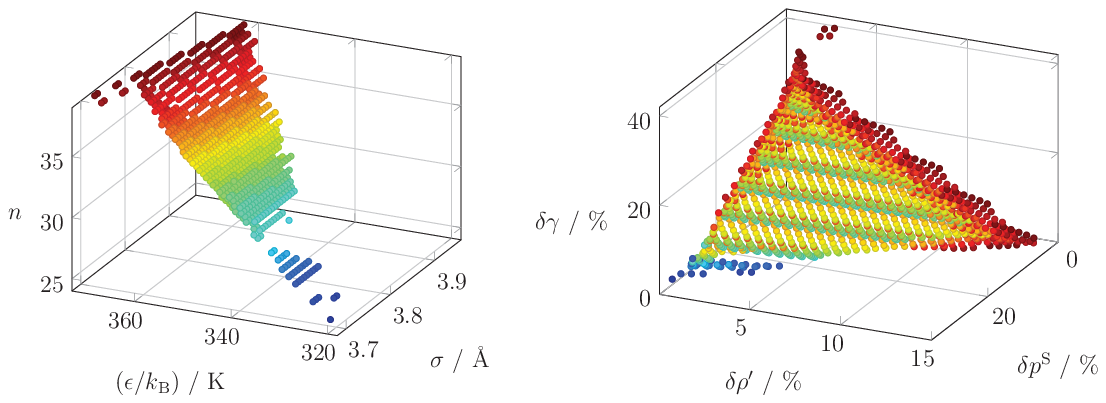}
 }
\caption{Pareto set of the Mie potential for CO$_2$ in the parameter space (left) and the objective space (right). %The symbols represent molecular models: Avenda\~{n}o et al. \cite{ALGAJM11} ($\triangle$) and the present model with $\sigma$ = 3.768 $\textup{\AA}$, $\epsilon$ / $k_{\rm B}$  = 366 K and $n$ = 39 (\Circle). 
The colors represent the numerical value of the repulsive exponent $n$ and connect the points in the parameter and the object space.}  
\label{fig:Mie_CO2}
\end{figure}

To reduce the deviations in the surface tension, the surface tension has to be included in the parameterization. Fig. \ref{fig:Mie_CO2} shows the three dimensional Pareto set of the single-site three-parameter Mie model for CO$_2$ in the parameter and objective spaces. The surface tension is added to the other two objective functions, i.e. the saturated liquid density and the vapor pressure. The Pareto set of the two-dimensional optimization is a subset of the Pareto set resulting from the three dimensional optimization. %The four-parameter Mie model by Avenda\~{n}o et al. \cite{ALGAJM11} is close to the three-dimensional Pareto set. Its accuracy is similar to what can be reached with the three-parameter Mie model over this set of objectives. The repulsive exponent $n$ varies between 24 and 39, which compares favorably with Avenda\~{n}o et al. \cite{ALGAJM11}, who used a repulsive exponent of $n$ = 23, but a different dispersive exponent, and Maurer \cite{M78}, who used exponents of $n$ = 30 and $n$ = 33 for 
%a perturbation theory. Ramrattan et al. concluded that a repulsive exponent $n=31$ should be used for an attractive exponent of $m=6$ \cite{RAMG15}.
Adding a third criterion to the optimization leads to a more complex situation. The surface tension is a competing objective function to the other objective functions, i.e. an optimization in the surface tension leads to a decline in at least one other objective function. It is usually not possible to obtain a good molecular model that represents the saturated liquid density, the vapor pressure and the surface tension with a good accuracy simultaneously \cite{LALMJ15,WSKKHH15,SKHKH15}.

\section{Conclusion}

In the present work the VLE of the Mie fluid was evaluated by molecular dynamics simulations. The LRC for molecular simulations with planar interfaces from previous work was generalized to the Mie potential. The influence of the LRC on the numerical accuracy for the saturated liquid density, the vapor pressure and the surface tension was studied. The present approach yields very good results, and its dependence on the cutoff radius is weak down to $r_{\rm c} = 2.5 \, \sigma$.

The VLE of the Mie fluid was determined with 14 different values of the repulsive exponent parameter, yielding results for the saturated liquid density, the saturated vapor density, the vapor pressure, the enthalpy of vaporization and the surface tension. A global correlation for the critical properties as well as the VLE properties was developed as a function of the repulsive exponent $n$. The correlations agree with the simulation data within the statistical uncertainties in most cases and are also in very good agreement with available simulation data on the Mie fluid from the literature.

Based on these correlations, new molecular models of the Mie type can easily be developed. As an example, CO$_2$ is studied. The parameterization of the Mie model of CO$_2$ is based on the correlations established in the present study. Multi-criteria optimization is used. The Pareto set gives an overview how well the Mie model can represent the studied properties, which are saturated liquid density, vapor pressure and surface tension. %the best compromises between multiple objectives for modeling CO$_2$ by a single Mie interaction site were determined and discussed. 
It is possible to obtain a molecular model that represents the saturated liquid density and the vapor pressure with 0.8 \% and 5.6 \% deviation, respectively. However, the average deviation of this model from the experimental surface tension is comparably high (23 \%). %The Pareto set determined in the present work characterizes how the accuracy for the bulk properties decreases as the accuracy for the surface tension is improved. 
These results can therefore be used for tuning the three-parameter Mie potential for CO$_2$ to individual needs. Furthermore, the correlations of the different properties enable a swift development of Mie models for other fluids.%, and similarly for other fluids also.

%%%%%%%%%%%%%%%%%%%%%%%%%%%%%%%%%%%%%%%%%%%%%%%%%%%%%%%%%%%%%%%%%%%%%
\section*{Acknowledgement}

The authors gratefully acknowledge financial support from BMBF within the SkaSim project (grant no. 01H13005A) and from Deutsche Forschungsgemeinschaft (DFG) within the Collaborative Research Center (SFB) 926. %They thank Carlos Avenda\~{n}o, Marco H\"ulsmann, Peter Klein, Karl-Heinz K\"ufer, Philippe Ungerer, Jadran Vrabec for fruitful discussions. 
\textbf{They greatly appreciate the advice from Herv\'e Gu\'erin who detected multiple errors in a previous version of the manuscript.}
The present work was conducted under the auspices of the Boltzmann-Zuse Society of Computational Molecular Engineering (BZS), and the simulations were carried out on the Regional University Computing Center Kaiserslautern (RHRK) under the grant TUKL-MSWS and on SuperMUC at Leibniz Supercomputing Center, Garching, within the SPARLAMPE scientific computing project.

%\end{acknowledgement}

%\appendix
\section*{Appendix}

\section*{Molecular simulation details}

The simulations were performed in the canonical ensemble. The equation of motion was solved by a leapfrog integrator~\cite{Fincham92} with a time step of $\Delta t$ = 0.001 $\sigma \sqrt{m / \epsilon}$. The elongation of the simulation volume normal to the interface was 80 $\sigma$ and the thickness of the liquid film in the center of the simulation volume was 40 $\sigma$ to account for finite size effects~\cite{WLHH13}. The elongation in the other spatial directions was at least 20 $\sigma$. 
The equilibration was executed for 500,000 time steps. The production was conducted for 2,500,000 time steps to reduce statistical uncertainties. Throughout the present work, the statistical errors were estimated to be three times the standard deviation of five block averages, each over 500,000 time steps. The saturated densities and vapor pressures were calculated as an average over the respective phases excluding the area close to the interface, i.e. the area where the first derivative of the density with respect to the $y$ coordinate deviated from zero significantly. 

\label{lastpage}

\bibliographystyle{tMPH}

\end{document}